\documentclass[aps,prd,onecolumn,preprintnumbers,groupedaddress,showpacs,nofootinbib,amssymb]{revtex4-2}
\usepackage{graphicx,color}
\usepackage{subcaption}
\usepackage{amsmath}
\usepackage{amssymb}
\usepackage{amsfonts}
\usepackage{bm}
\usepackage{cancel}

\newcommand{\e}{\mathrm{e}}

\allowdisplaybreaks[4] 
\begin{document}

\preprint{KEK-TH-2774, KEK-Cosmo-0396}
\title{
Wormhole spacetimes in an expanding universe: \\
Energy conditions and future singularities
}

\author{Taishi~Katsuragawa$^1$}\email{taishi@ccnu.edu.cn}
\author{Shin'ichi~Nojiri$^{2,3}$}\email{nojiri@nagoya-u.jp} 
\author{Sergei~D.~Odintsov$^{4,5}$}\email{odintsov@ice.csic.es}
\affiliation{
1. Institute of Astrophysics, Central China Normal University, Wuhan 430079, China \\
2. KEK Theory Center, High Energy Accelerator Research Organization (KEK), Oho 1-1, Tsukuba, Ibaraki 305-0801, Japan \\
3. Kobayashi-Maskawa Institute for the Origin of Particles and the Universe, 
4. ICREA, Passeig Luis Companys, 23, 08010 Barcelona, Spain \\
5. Institute of Space Sciences (ICE, CSIC) C. Can Magrans s/n, 08193 Barcelona, Spain
}

\begin{abstract}
We study wormhole geometries embedded in an expanding universe within a four-scalar non-linear $\sigma$ model, where the target-space metric is identified with the spacetime Ricci tensor. 
In this framework, wormholes can remain stable even when conventional energy conditions are violated. 
However, once cosmological expansion is included, the effective energy density and pressure are modified by the cosmological fluid, enabling the energy conditions to be satisfied. 
We further present intriguing geometries in which a finite future singularity appears in our universe but not in another universe connected by the wormhole. 
Near the throat, the hypersurface becomes timelike, allowing trajectories to traverse to the other universe before the singularity and return afterwards.
We also construct wormhole solutions motivated by galactic dark-matter halo profiles, where the required non-vanishing pressure arises naturally from the four-scalar non-linear $\sigma$ model.
\end{abstract}

\maketitle

\section{Introduction}
\label{SEcI}

An exotic compact object called a wormhole was first proposed in Refs.~\cite{Misner:1960zz, Wheeler:1957mu}.
Such geometries connect distant regions of the universe or even entirely separate universes~\cite{Morris:1988cz}.
These structures require matter that violates the null energy condition (NEC) and, hence, all the standard energy conditions, including the weak energy condition (WEC), strong energy condition (SEC), and dominant energy condition (DEC).
Such matter is often referred to as exotic matter and typically leads to instabilities.
Possible origins of such matter include higher-dimensional effects~\cite{Bronnikov:2002rn}.
Wormholes have attracted renewed attention not only as theoretical curiosities but also as potential probes of fundamental physics.

A pioneering work~\cite{Maeda:2009tk} showed the construction of the wormhole solution embedded in the expanding universe satisfying the energy conditions.
Moreover, effects of the non-local gravity, the inverse d'Alembert operator in the gravitational sector, can stabilize wormhole solutions~\cite{Capozziello:2022zoz}, and the wormhole spacetime can be supported without exotic matter in the Einstein-Gauss-Bonnet gravity with the four-dimensional limit~\cite{Godani:2024rqf}.
Building on numerous earlier results, we aim to investigate the concrete realization of the energy–momentum tensor required to support the wormhole geometry within a self-consistent matter sector.
We also demonstrate a novel field-theoretical framework in which the wormhole solution is supported, rather than relying on an effective or phenomenological fluid description.

In this work, we study wormholes embedded in an expanding universe.
Although phantom dark energy (DE) violates all the energy conditions, the presence of the cosmological fluid responsible for the expansion can modify the total energy density and pressure, thereby relaxing the violation, as discussed in Ref.~\cite{Bronnikov:2023saq}.
A related model was proposed in Ref.~\cite{Nojiri:2023dvf}, where Einstein gravity coupled to two scalar fields generates a wormhole geometry.
In this model, the energy conditions are violated due to the energy–momentum tensor of the scalar fields, and one of the fields becomes a ghost.
Classically, such a ghost leads to an unbounded kinetic energy, and in quantum theory, it gives rise to negative-norm states, which undermine the consistency of quantum mechanics.
However, the ghost can be removed by introducing a Lagrange multiplier field, which imposes a constraint that effectively freezes the scalar dynamics, leading to a non-dynamical but stable configuration.

An extension including a Gauss–Bonnet term was proposed in Ref.~\cite{Elizalde:2023rds}, leading to time-dependent, or dynamical, wormhole geometries that exist only within a finite time interval.
To construct a general spherically symmetric and dynamical geometry, two scalar fields are sufficient.
However, to realise an entirely arbitrary geometry, four scalar fields are necessary and sufficient~\cite{Nashed:2024jqw, Katsuragawa:2024bwm}.
In this framework, the dynamical wormhole geometries are formulated as a non-linear $\sigma$ model in which the target-space metric is identified with the Ricci tensor of the spacetime~\cite{Nojiri:2025heh}, and the four scalar fields correspond to spacetime coordinates.
Because of the inherent constraints in this model, the geometry is stable even when the energy conditions are violated.
The scalar fields and non-linear $\sigma$ models often appear in the theories originated from higher-dimensional theories like string theory,
and the gauge symmetries, including general covariance in higher dimensions, may give constraints as in the non-linear $\sigma$ model.

Using this four-scalar framework, we can construct a broad class of wormhole geometries.
First, we examine a wormhole embedded in an expanding universe.
Although the configuration remains stable despite violating the energy conditions, we demonstrate that cosmological expansion can shift the total energy density and pressure so that the conditions are effectively satisfied.
We also consider more exotic configurations in which our universe encounters a finite future singularity (see \cite{deHaro:2023lbq} as a review), while another universe connected via the wormhole remains non-singular.
Near the wormhole throat, the singular hypersurface becomes timelike, allowing a possible timelike trajectory in which a particle can traverse to the other universe before the singularity and return afterwards.

Finally, we investigate wormhole geometries motivated by galactic dark-matter halo profiles, such as the Navarro–Frenk–White (NFW) profile~\cite{Navarro:1995iw, Navarro:1996gj} and related models~\cite{Begeman:1991iy, Boehmer:2007um, Errehymy:2025vvs}.
We show that in such configurations the pressure cannot vanish, implying that the supporting matter cannot be ordinary dark matter but rather an exotic component, which can naturally be realised within the four-scalar non-linear $\sigma$ model.

The structure of the paper is as follows.
Sec.~\ref{SecII} reviews the non-linear $\sigma$ model that realises arbitrary geometries.
Sec.~\ref{SecIII} presents dynamical wormhole solutions based on the simplest static case.
Sec.~\ref{SecIV} summarises the standard energy conditions and demonstrate their violation in the simplest wormhole geometry.
Sec.~\ref{SecV} reviews the classification of future singularities in the Friedmann–Lemaître–Robertson–Walker (FLRW) universe.
Sec.~\ref{SecVI} presents the embedding of a wormhole in an expanding universe.
Sec.~\ref{SecVII} illustrates how the energy conditions can be satisfied or violated.
Sec.~\ref{SecVIII} discusses wormhole configurations connecting a singular and a non-singular universe.
Sec.~\ref{SecIX} is devoted to wormholes inspired by dark-matter halo profiles, and Sec.~\ref{SecX} concludes with summary and discussion.

\section{Spacetime and target-space geometries in non-linear $\sigma$ model with four scalar fields}
\label{SecII}

In this section, we show that an arbitrary geometry can be realised using a non-linear $\sigma$ model whose target-space metric is identified with the Ricci curvature~\cite{Nashed:2024jqw, Katsuragawa:2024bwm}.
In four dimensions, the model contains four scalar fields corresponding to the spacetime coordinates.
It should be noted that the four-scalar model is the minimal setup capable of realizing any geometry in four dimensions in a stable manner.
More generally, in $d$ dimensions, the minimal model requires $d$ scalar fields.

We consider the following non-linear $\sigma$ model, which includes four scalar fields $\phi^{a}$ $(a=0,1,2,3)$ and is coupled to Einstein gravity,
\begin{align}
\label{Eq. acg1}
    S 
    &= 
    S_\mathrm{gravity} + S_\phi + S_\lambda 
    \, , \\
\label{Einstein} 
    S_\mathrm{gravity} 
    &= 
    \frac{1}{2\kappa^2} \int d^4 x \sqrt{-g} R 
    \, , \\
\label{Eq. acg2}
    S_\phi 
    &\equiv 
    \frac{1}{2} \int d^4x \sqrt{-g} \sum_{a,b = 0,1,2,3} A_{ab} \left( \phi \right) g^{\mu\nu} \partial_\mu \phi^{a} \partial_\nu \phi^{b} 
    \, , \\
\label{Eq. acg3}
    S_\lambda 
    &\equiv
    \int d^4x \sqrt{-g} \sum_{a=0,1,2,3} \lambda^{a} \left( \frac{1}{g^{aa}\left( x = \phi \right)} g^{\mu\nu} 
    \left( x\right) \partial_\mu \phi^{a} \partial_\nu \phi^{a} - 1 \right) 
    \, .
\end{align}
Here, we use Roman indices ($a, b, \dots = 0,1,2,3$) for the scalar fields, corresponding to indices in the internal space.
The kinetic coefficients $A_{ab}(\phi)$ are functions of the scalar fields $\phi^a$.
In Eq.~\eqref{Eq. acg3}, $\lambda^a$'s are Lagrange multiplier fields that impose the following constraints:
\begin{align}
\label{Eq: cnstrnt1}
    0 
    = 
    \frac{1}{g^{aa} \left( x = \phi \right)} g^{\mu\nu} \left( x \right) \partial_\mu \phi^{a} \partial_\nu \phi^{a} - 1 
    \, .
\end{align}
Due to the constraints in Eq.~\eqref{Eq: cnstrnt1}, ghosts can be eliminated~\cite{Nashed:2024jqw, Katsuragawa:2024bwm}.
Furthermore, these constraints prevent the scalar fields from propagating.
Although the scalar fields mimic a fluid in terms of energy density and pressure, they are not associated with sound propagation.
Consequently, even if any energy condition is violated, no instability arises from the scalar fields.

By the variation of the action \eqref{Eq. acg1} with respect to the metric $g_{\mu\nu}$, we obtain
\begin{align}
\label{Eq: Eqs1}
    \mathcal{G}_{\mu\nu} 
    &=
    \frac{1}{4} g_{\mu\nu} \sum_{a, b=0,1,2,3} A_{ab} \left(\phi \right) 
    - \frac{1}{2} \sum_{a,b = 0,1,2,3} A_{ab} \left(\phi\right) \partial_\mu \phi^{a} \partial_\nu \phi^{b} 
    \nonumber \\
    & \quad 
    + \frac{1}{2} g_{\mu\nu} \sum_{a=0,1,2,3} \lambda^{a} \left( \frac{1}{g^{aa}\left( x = \phi \right)} g^{\mu\nu} \left( x \right) \partial_\mu \phi^{a} \partial_\nu \phi^{a} - 1 \right) 
    - \sum_{a=0,1,2,3} \frac{\lambda^{a}}{g^{aa} \left( x = \phi \right)} \partial_\mu \phi^{a} \partial_\nu \phi^{a} 
    \nonumber \\
    &= 
    \frac{1}{4} g_{\mu\nu} \sum_{a, b=0,1,2,3} A_{ab} \left(\phi \right) g^{\xi\eta} \partial_\xi \phi^{a} \partial_\eta \phi^{b} 
    - \frac{1}{2} \sum_{a,b=0,1,2,3} A_{ab} \left(\phi \right) \partial_\mu \phi^{a} \partial_\nu \phi^{b} 
    - \sum_{a=0,1,2,3} \frac{\lambda^{a}}{g^{aa} \left( x = \phi \right)} \partial_\mu \phi^{a} \partial_\nu \phi^{a} 
    \, .
\end{align}
Here, we used the constraint equations in Eq.~\eqref{Eq: cnstrnt1}, 
and $\mathcal{G}_{\mu\nu}$ is given by the Einstein tensor $G_{\mu\nu}$, 
\begin{align}
\label{Eq: GR case}
\mathcal{G}_{\mu\nu} = - \frac{1}{2\kappa^2} G_{\mu\nu} = - \frac{1}{2\kappa^2} \left( R_{\mu\nu} - \frac{1}{2} g_{\mu\nu} R \right) \, .
\end{align}
In this paper, we neglect the contribution from matter.  

Contracting Eq.~\eqref{Eq: Eqs1} with the metric $g^{\mu\nu}$ yields
\begin{align}
\label{Eqs2}
    g^{\mu\nu} \mathcal{G}_{\mu\nu} 
    = 
    \frac{1}{2} \sum_{a,b=0,1,2,3} A_{ab} \left(\phi \right) g^{\xi\eta} \partial_\xi \phi^{a} \partial_\eta \phi^{b} 
    - \sum_{a=0,1,2,3} \lambda^{a} 
    \, ,
\end{align}
where we again used Eq.~\eqref{Eq: cnstrnt1}. 
Furthermore, by substituting Eq.~\eqref{Eqs2} into Eq.~\eqref{Eq: Eqs1}, we find
\begin{align}
\label{Eqs3}
    \sum_{a,b=0,1,2,3} A_{ab} \left(\phi\right) \partial_\mu \phi^{a} \partial_\nu \phi^{b}
    = 
    - 2 \mathcal{G}_{\mu\nu} + g_{\mu\nu} \left\{ \sum_{a=0,1,2,3} \lambda^{a} + g^{\rho\sigma} \mathcal{G}_{\rho\sigma} \right\} 
    - 2 \sum_{a=0,1,2,3} \frac{\lambda^{a}}{g^{aa}\left( x = \phi \right)} \partial_\mu \phi^{a} \partial_\nu \phi^{a} 
    \, .
\end{align}
The four scalar fields are identified with the spacetime coordinates $\phi^{a}=x^a$, which is consistent with the constraints in Eq.~\eqref{Eq: cnstrnt1}. 
Using this identification, we rewrite Eq.~\eqref{Eqs3} as follows, 
\begin{align}
\label{Eqs4}
A_{\mu\nu} \left(\phi \right) = - 2 \mathcal{G}_{\mu\nu} + g_{\mu\nu} \left\{ \sum_{a=0,1,2,3} \lambda^{a} + g^{\rho\sigma} \mathcal{G}_{\rho\sigma} \right\}
 - 2 \sum_{a=0,1,2,3} \frac{\lambda^{a}}{g^{aa}\left( x = \phi \right)} \delta_{\mu}^{a} \delta_{\nu}^{a} \, .
\end{align}
Here, we may consider the solution $\lambda^a = 0$.
This indicates that an arbitrary geometry specified by the metric $g_{\mu\nu}$ can be realised by choosing the kinetic coefficients $A_{\mu\nu}(\phi)$ as
\begin{align}
\label{Eqs5}
    A_{\mu\nu} \left(\phi \right) 
    = 
    - 2 \mathcal{G}_{\mu\nu} \left( x = \phi \right) 
    + g_{\mu\nu} \left( x = \phi \right) 
    g^{\rho\sigma} \left( x = \phi \right) 
    \mathcal{G}_{\rho\sigma} \left( x = \phi \right) 
    \, .
\end{align}
The action $S_{\phi}$ can be regarded as describing a non-linear $\sigma$ model whose target-space metric is given by $A_{ab}(\phi)$.
When $A_{ab} = 0$ for a given $a$ and arbitrary $b$, and the remaining non-vanishing components do not depend on $\phi^a$ for that $a$, the scalar field $\phi^a$ can be removed.
For example, we consider the spherically symmetric spacetime.
In this spacetime, the scalar fields associated with the angular coordinates, $\phi^2 = \theta$ and $\phi^3 = \varphi$, do not appear in the action.
This implies that two of the four scalar fields can be removed, and a two-scalar model suffices to describe the spherically symmetric spacetime~\cite{Nojiri:2020blr}.

In the case of Einstein gravity, Eq.~\eqref{Eqs5} is explicitly given by,
\begin{align}
\label{Eqs5B}
    A_{\mu\nu} \left(\phi \right) 
    = 
    \frac{1}{\kappa^2} G_{\mu\nu} \left( x = \phi \right) 
    - \frac{1}{2\kappa^2} g_{\mu\nu} \left( x = \phi \right) g^{\rho\sigma} \left( x = \phi \right) G_{\rho\sigma} \left( x = \phi \right) 
    = 
    \frac{1}{\kappa^2} R_{\mu\nu} \left( x = \phi \right) \, .
\end{align}
Therefore, $A_{\mu\nu} \left(\phi \right)$ is given by the Ricci tensor $R_{\mu\nu}$, where the coordinates are identified with the scalar fields $x^{\mu} = \phi^{\mu}$. 
Arbitrary geometry can be realised by a non-linear $\sigma$ model whose target space metric is given by the Ricci curvature.

\section{Dynamical wormhole spacetime}
\label{SecIII}

In this section, we propose time-dependent, or dynamical, wormhole geometries.
We begin by considering the simplest wormhole geometry, given by
\begin{align}
\label{wh1} 
    ds^2 
    = 
    - \e^{2\nu} dt^2 + \e^{2\lambda} dr^2 
    + r^2 \left( d\theta^2 + \sin^2\theta d\phi^2 \right) 
    \, , \quad 
    \e^{2\nu} = 1 
    \, , \quad 
    \e^{2\lambda}=\frac{1}{1 - \frac{r_0}{r}} 
    \, ,
\end{align}
where $r_0$ is the minimum radius of the wormhole. 
When $r \sim r_0$, we have $e^{2\lambda} \sim \frac{r_0}{r - r_0}$, which appears singular at $r = r_0$.
This singularity, however, is not physical and can be removed by a coordinate transformation.
To eliminate the apparent singularity, we introduce a new coordinate $\chi$ defined as follows:
\begin{align}
\label{lll}
    r = \sqrt{ {r_0}^2 + \chi^2} 
    \, ,\quad 
    dr= \frac{\chi d\chi}{\sqrt{ {r_0}^2 + \chi^2}} 
    \, .
\end{align}
Eq.~\eqref{wh1} is transformed into the following form,
\begin{align}
\label{lll2}
    ds^2 
    = 
    - dt^2 + \frac{\chi^2 d\chi^2}{{r_0}^2 + \chi^2 - r_0 \sqrt{ {r_0}^2 + \chi^2}} 
    + \left(  {r_0}^2 + \chi^2 \right) \left( d\theta^2 + \sin^2\theta d\phi^2 \right)
    \, . 
\end{align}
When $r \to r_0$ in the original coordinate, we have $\chi \to 0$, and
\begin{align}
    \frac{\chi^2}{{r_0}^2 + \chi^2- r_0 \sqrt{ {r_0}^2 + \chi^2}} \to 2
    \, .
\end{align} 
Thus, in the new coordinate system, the spacetime is regular at $r = r_0$ ($\chi = 0$).
We analytically continue $\chi$ into the region where it takes negative values, $-\infty < \chi<+ \infty$.
If the region with positive $\chi$ corresponds to our Universe, the region with negative $\chi$ corresponds to another Universe.
The two Universes are connected by a wormhole, whose minimum radius is $r_0$.
We also note that in the limit $|\chi| \to \infty$, we have $|\chi| \sim r$ and
\begin{align}
    \frac{\chi^2}{{r_0}^2 + \chi^2 - r_0 \sqrt{ {r_0}^2 + \chi^2}} \to 1
    \, ,
\end{align}
so that the spacetime is asymptotically flat.
For the geometry in Eqs.~\eqref{wh1} and \eqref{lll2}, we find that $R = 0$.

We now consider the following dynamical geometry instead of Eq.~\eqref{lll2}: 
\begin{align}
\label{TD1}
    ds^2 
    = 
    \e^{2N(\tau,\chi)} 
    \left\{
    - d\tau^2 + \frac{\chi^2 d\chi^2}{{r_0}^2 + \chi^2 - r_0 \sqrt{ {r_0}^2 + l^2}} 
    + \left(  {r_0}^2 + \chi^2 \right) \left( d\theta^2 + \sin^2\theta d\phi^2 \right) 
    \right\}
    \, .
\end{align}
Then the non-vanishing components of the Christoffel symbols are given by 
\begin{align}
\label{TD2}
    \Gamma^{\tau}_{\tau\tau} 
    &= 
    N^{(1,0)}(\tau,\chi) 
    \, , \quad 
    \Gamma^{\tau}_{\chi\tau} 
    = 
    N^{(0,1)}(\tau,\chi) 
    \, , \quad 
    \Gamma^{\tau}_{\chi\chi} 
    =
    \left(\frac{r_0}{\sqrt{{r_0}^2+\chi^2}}+1\right) N^{(1,0)}(\tau,\chi) 
    \, , \nonumber \\
    \Gamma^{\tau}_{\theta \theta} 
    &= 
    \left({r_0}^2+\chi^2\right) N^{(1,0)}(\tau,\chi) 
    \, , \quad 
    \Gamma^{\tau}_{\phi\phi} 
    = 
    \sin^2 \theta \left({r_0}^2+\chi^2\right) N^{(1,0)}(\tau,\chi) 
    \, , \nonumber \\
    \Gamma^{\chi}_{\tau\tau} 
    &= 
    \frac{{r_0}^2+\chi^2-r_0 \sqrt{{r_0}^2+\chi^2} }{\chi^2} N^{(0,1)}(\tau,\chi) 
    \, , \quad 
    \Gamma^{\chi}_{\chi\tau} 
    = 
    \Gamma^{\chi}_{\tau\chi} = N^{(1,0)}(\tau,\chi) 
    \, , \nonumber \\
    \Gamma^{\chi}_{\chi\chi} 
    &=
    N^{(0,1)}(\tau,\chi)+\frac{{r_0}^2-r_0 \sqrt{{r_0}^2+\chi^2}}{2\chi({r_0}^2 + \chi^2)} 
    \, , \nonumber \\
    \Gamma^{\chi}_{\theta\theta} 
    &=
    -\frac{1}{\chi^2} \left({r_0}^2+\chi^2-r_0 \sqrt{{r_0}^2+\chi^2}\right) 
    \left[ \left({r_0}^2+\chi^2\right) N^{(0,1)}(\tau,\chi)+\chi\right] 
    \, , \nonumber \\
    \Gamma^{\chi}_{\phi\phi} 
    &=
    - \frac{\sin^2 \theta}{\chi^2} \left({r_0}^2+\chi^2-r_0 \sqrt{{r_0}^2+\chi^2}\right) 
    \left[\left({r_0}^2+\chi^2\right) N^{(0,1)}(\tau,\chi)+\chi \right] 
    \, , \nonumber \\
    \Gamma^{\theta}_{\theta \tau} 
    &= 
    N^{(1,0)}(\tau,\chi) 
    \, , \quad 
    \Gamma^{\theta}_{\theta \chi} 
    = 
    N^{(0,1)}(\tau,\chi)+\frac{\chi}{{r_0}^2+\chi^2} 
    \, , \quad 
    \Gamma^{\theta}_{\phi\phi} 
    = 
    - \sin \theta \cos \theta 
    \, , \nonumber \\
    \Gamma^{\phi}_{\phi \tau} 
    &=
    N^{(1,0)}(\tau,\chi) 
    \, , \quad 
    \Gamma^{\phi}_{\phi \chi} 
    = 
    N^{(0,1)}(\tau,\chi)+\frac{\chi}{{r_0}^2+\chi^2} 
    \, , \quad 
    \Gamma^{\phi}_{\phi\theta} 
    = 
    \cot \theta 
    \, .
\end{align}
Here $N^{(n,m)}$ denotes $n$th derivative with respect to $\tau$ and $m$th derivative with respect to $\chi$, $N^{(n,m)} \equiv \frac{\partial^{n+m} N(\tau,\chi)}{\partial^n \tau \partial^m \chi}$. 

In this paper, we use the following definitions of curvatures, 
\begin{align}
\label{curvaturesB}
    R^{\lambda}_{\ \mu\rho\nu} 
    &= 
    \Gamma^\lambda_{\mu\nu,\rho} -\Gamma^\lambda_{\mu\rho,\nu} + \Gamma^\eta_{\mu\nu}\Gamma^\lambda_{\rho\eta}
    - \Gamma^\eta_{\mu\rho}\Gamma^\lambda_{\nu\eta} 
    \, , \nonumber \\
    R_{\mu\nu} 
    &=
    {R^\rho}_{\ \mu\rho\nu} = \Gamma^\rho_{\mu\nu,\rho} -\Gamma^\rho_{\mu\rho,\nu} + \Gamma^\eta_{\mu\nu}\Gamma^\rho_{\rho\eta}
    - \Gamma^\eta_{\mu\rho}\Gamma^\rho_{\nu\eta} 
    \, , \nonumber \\
    R 
    &=
    g^{\mu\nu} R_{\mu\nu} 
    \, .
\end{align}
Then, we obtain the non-vanishing components of the Ricci tensor and the Ricci scalar as follows:
\begin{align}
\label{TD3}
    R_{\tau \tau} 
    &= 
    -3 N^{(2,0)}(\tau, \chi)
    - \frac{{r_0}^2+\chi^2}{\chi^2} \left( \frac{r_0}{\sqrt{{r_0}^2+\chi^2}} - 1 \right) N^{(0,2)}(\tau, \chi) 
    \nonumber \\
    & \qquad 
    - 2\frac{{r_0}^2+\chi^2}{\chi^2} \left( \frac{r_0}{\sqrt{{r_0}^2+\chi^2}} - 1 \right) N^{(0,1)}(\tau, \chi)^2 
    + \frac{1}{\chi^3}\left[ \left( 2\chi^2-{r_0}^2 \right) + \frac{r_0(2{r_0}^2-3\chi^2)}{2\sqrt{{r_0}^2+\chi^2}} \right] N^{(0,1)}(\tau, \chi) 
    \, , \nonumber \\
    R_{\tau \chi} 
    &= 
    2 N^{(0,1)}(\tau, \chi) N^{(1,0)}(\tau, \chi)-2 N^{(1,1)}(\tau, \chi) 
    \, , \nonumber \\
    R_{\chi \chi} 
    &=
    \left( \frac{r_0}{\sqrt{{r_0}^2+\chi^2}} + 1\right) N^{(2,0)}(\tau, \chi) 
    -3 N^{(0,2)}(\tau, \chi)
    + 2 \left( \frac{r_0}{\sqrt{{r_0}^2+\chi^2}} + 1\right) N^{(1,0)}(\tau, \chi)^2 
    \nonumber \\
    & \qquad 
    + \frac{ \left( 3{r_0}^2 -4\chi^2\right) -3 r_0\sqrt{{r_0}^2+\chi^2}}{2\chi\left({r_0}^2+\chi^2\right)} N^{(0,1)}(\tau, \chi) 
    - \frac{r_0}{\left({r_0}^2+\chi^2\right)^{3/2}} \left( \frac{r_0}{\sqrt{{r_0}^2+\chi^2}} + 1\right) 
    \, , \nonumber \\
    R_{\theta \theta} 
    &=
    \left({r_0}^2+\chi^2\right) N^{(2,0)}(\tau, \chi) 
    + \frac{\left({r_0}^2+\chi^2\right)^{2}}{\chi^2} \left( \frac{r_0}{\sqrt{r^2+\chi^2}} -1\right) N^{(0,2)}(\tau, \chi) 
    \nonumber \\
    & \qquad 
    + 2\left({r_0}^2+\chi^2\right) N^{(1,0)}(\tau, \chi)^2
    + \frac{2\left({r_0}^2+\chi^2\right)^{2} }{\chi^2} \left( \frac{r_0}{\sqrt{{r_0}^2+\chi^2}} -1 \right) N^{(0,1)}(\tau, \chi)^2 
    \nonumber \\
    &\qquad 
    + \frac{{r_0}^2+\chi^2}{\chi^3} \left[ \left({r_0}^2 - 4\chi^2 \right) 
    + \frac{r_0 \left(7  \chi^2 -2 {r_0}^2 \right)}{2\sqrt{{r_0}^2+\chi^2}} \right] N^{(0,1)}(\tau, \chi)
    + \frac{r_0}{2\sqrt{{r_0}^2+\chi^2}} 
    \, , \nonumber \\
    R_{\phi\phi} 
    &=
    R_{\theta \theta} \sin^2 \theta 
    \, , \nonumber \\
    R 
    &=
    \e^{-2 N(\tau, \chi)} 
    \left\{ 
    6 N^{(2,0)}(\tau, \chi) 
    + \frac{6\left({r_0}^2+\chi^2\right)}{\chi^2} \left( \frac{r_0}{\sqrt{{r_0}^2+\chi^2}} - 1\right) N^{(0,2)}(\tau, \chi) 
    +6 N^{(1,0)}(\tau, \chi)^2 
    \right.
    \nonumber \\
    &\qquad \qquad \qquad
    + \frac{6 \left( {r_0}^2+\chi^2\right)}{\chi^2} \left( \frac{r_0}{\sqrt{{r_0}^2+\chi^2}} -1 \right) N^{(0,1)}(\tau, \chi)^2 
    \nonumber \\
    &\qquad \qquad \qquad
    \left.
    +\frac{3}{\chi^3} \left[ 2 \left( {r_0}^2 -2 \chi^2 \right) + \frac{r_0\left( 3  \chi^2-2 {r_0}^2 \right)}{\sqrt{{r_0}^2+\chi^2}}
    \right] N^{(0,1)}(\tau, \chi) 
    \right\}
    \, .
\end{align}

Replacing the coordinates $(\tau, \chi, \theta, \phi)$ with four scalar fields $(\phi_0, \phi_1, \phi_2, \phi_3)$ and using Eq.~\eqref{Eqs5B},
we obtain the corresponding non-linear $\sigma$ model in Eq.~\eqref{Eq. acg2} that realises the dynamical wormhole given in Eq.~\eqref{TD1}.
If $N(\tau, \chi)$ is chosen as a smooth function of $\tau$ and $\chi$, a time-dependent, or dynamical, wormhole solution can be constructed, as in Ref.~\cite{Alencar:2025nik}.
We also note that a non-trivial value of $R_{\tau\chi}$ corresponds to the radial energy flux, $j = \frac{1}{\kappa^2} R_{\tau\chi}$.

\section{Energy conditions}
\label{SecIV}

In this section, we introduce four conventional energy conditions and formulate these conditions for the wormhole solution obtained in the previous section. 
Using Eq.~\eqref{TD1}, we demonstrate that all the energy conditions are violated in the simplest case $N=0$.
For NEC, WEC, SEC and DEC, the energy-momentum tensor $T_{\mu\nu}$ satisfies the following inequalities, respectively: 
\begin{align}
    \text{NEC}:& \quad 
    T_{\mu\nu} k^\mu k^\nu \geq 0
    \, , \nonumber \\
    \text{WEC}:& \quad 
    T_{\mu\nu} V^\mu V^\nu \geq 0
    \, ,\nonumber \\
    \text{SEC}:& \quad 
    \left(T_{\mu\nu}-\frac{1}{2}g_{\mu\nu}T \right)V^\mu V^\nu \geq 0
    \, ,\nonumber \\
    \text{DEC}:& \quad 
    T_{\mu\nu} V^\mu V^\nu \geq 0
    \, , \ \text{and} \ T_{\mu\nu} V^\nu \ \text{is not spacelike}
    \, . 
\label{Eq: energy cond1}
\end{align}
for any null vector $k^\mu$, $g_{\mu\nu}k^\mu k^\nu =0$, and for any time-like vector $V^\mu$, $g_{\mu\nu}V^\mu V^\nu <0$. 
For the wormhole geometry under our consideration, we have the energy-momentum tensor $T^\mu_{\ \nu}=\text{diag}\left(-\rho , p^\mathrm{radial}, p^\mathrm{angular}, p^\mathrm{angular}\right)$.  
Here $\rho$, $p^\mathrm{radial}$, $p^\mathrm{angular}$ are the rest-mass energy density and pressures in the radial direction and the angular direction, respectively. 
In terms of these quantities, Eq.~\eqref{Eq: energy cond1} yield 
\begin{align}
    \text{NEC}:& \quad 
    \rho + p^\mathrm{radial} \geq 0
    \, , \quad 
    \rho + p^\mathrm{angular} \geq 0
    \, , \nonumber \\
    \text{WEC}:& \quad 
    \rho + p^\mathrm{radial} \geq 0
    \, , \quad 
    \rho + p^\mathrm{angular} \geq 0
    \, , \quad 
    \rho \geq 0
    \, ,\nonumber \\
    \text{SEC}:& \quad 
    \rho + p^\mathrm{radial} \geq 0
    \, , \quad 
    \rho + p^\mathrm{angular} \geq 0
    \, , \quad 
    \rho +p^\mathrm{radial} +2 p^\mathrm{angular}\geq 0
    \, ,\nonumber \\
    \text{DEC}:& \quad 
    \rho \geq 0
    \, , \quad 
    \rho \geq | p^\mathrm{radial}|
    \, , \quad 
    \rho \geq | p^\mathrm{angular}| 
    \, .
\label{Eq: energy cond2}
\end{align}
Usually, a wormhole spacetime violates some of the energy conditions,
and we consider the energy conditions for the spacetime of the dynamical wormhole in Eq.~\eqref{TD1}. 

Using the Ricci tensor and Ricci scalar in Eq.~\eqref{TD3}, we obtain the non-vanishing components of the Einstein tensor for the geometry defined in Eq.~\eqref{TD1} as follows:
\begin{align}
    \label{NEinstein}
    G_{\tau \tau} 
    &=
    \, \frac{2 \left( {r_0}^2+\chi^2 \right)}{\chi^2} \left( \frac{r_0}{\sqrt{{r_0}^2+\chi^2}} - 1 \right) N^{(0,2)}(\tau, \chi) 
    +3 N^{(1,0)}(\tau, \chi)^2 
    \nonumber \\
    &
    + \frac{{r_0}^2+\chi^2}{\chi^2} \left( \frac{r_0}{\sqrt{{r_0}^2+\chi^2}} - 1 \right) N^{(0,1)}(\tau, \chi)^2 
     - \frac{1}{\chi^3} \left[ 2 \left( 2 \chi^2- {r_0}^2 \right) +\frac{ r_0(2{r_0}^2- 3\chi^2) }{\sqrt{{r_0}^2+\chi^2}} 
    \right] N^{(0,1)}(\tau, \chi) 
    \, , \nonumber \\
    G_{\tau \chi} 
    &=
    2 N^{(0,1)}(\tau, \chi) N^{(1,0)}(\tau, \chi)-2 N^{(1,1)}(\tau, \chi) 
    \, , \nonumber \\
    G_{\chi \chi} 
    &=
    - 2 \left( \frac{r_0}{\sqrt{{r_0}^2+\chi^2}} + 1\right) N^{(2,0)}(\tau, \chi) 
     - \left( \frac{r_0}{\sqrt{{r_0}^2+\chi^2}} + 1 \right) N^{(1,0)}(\tau, \chi)^2 
     \nonumber \\
    &
    +3 N^{(0,1)}(\tau, \chi)^2 
    +\frac{4\chi}{{r_0}^2+\chi^2} N^{(0,1)}(\tau, \chi)
     - \frac{r_0}{\left({r_0}^2+\chi^2\right)^{3/2}} \left( \frac{r_0}{\sqrt{{r_0}^2+\chi^2}} + 1 \right) 
     \, , \nonumber \\
    G_{\theta \theta} 
    &=
    -2 \left({r_0}^2+\chi^2\right) N^{(2,0)}(\tau, \chi)
     - \frac{2 \left({r_0}^2+\chi^2\right)^2}{\chi^2} \left( \frac{r_0}{\sqrt{{r_0}^2+\chi^2}} -    1 \right) N^{(0,2)}(\tau, \chi) 
     \nonumber \\
    &
    - \left({r_0}^2+\chi^2\right) N^{(1,0)}(\tau, \chi)^2
     - \frac{\left({r_0}^2+\chi^2\right)^2}{\chi^2} \left( \frac{r_0}{\sqrt{{r_0}^2+\chi^2}} - 1 \right) N^{(0,1)}(\tau, \chi)^2 
     \nonumber \\
    &
    + \frac{{r_0}^2 + \chi^2}{\chi^3} \left[ 2(\chi^2 - {r_0}^2) + \frac{r_0(2{r_0}^2-\chi^2)}{\sqrt{{r_0}^2+\chi^2}}
    \right] N^{(0,1)}(\tau, \chi)
    + \frac{r_0}{2 \sqrt{{r_0}^2+\chi^2} } 
    \, , \nonumber \\
    G_{\phi\phi} 
    &=
    \sin^2 \theta   G_{\theta \theta} 
    \, .
\end{align}
Moreover, using the Einstein equation, 
\begin{align}
\label{Ee}
    G_{\mu\nu} \equiv R_{\mu\nu} - \frac{1}{2} g_{\mu\nu} R = \kappa^2 T_{\mu\nu}
    \, , 
\end{align}
with a gravitational coupling $\kappa$, we can deduce the energy-momentum tensor; that is, the rest-mass energy density, radial and angular pressures, and energy flux, which realises the geometry given in Eq.~\eqref{TD1}. 

As an illustration, we examine the $N=0$ case, which corresponds to Eq.~\eqref{wh1}.
Substituting $N=0$ into Eq.~\eqref{NEinstein}, we obtain 
\begin{align}
\label{N0Einstein}
    G_{\tau\tau} = G_{\tau\chi} = 0
    \, , \quad 
    G_{\chi\chi} = - \frac{{r_0}^2 + r_0 \sqrt{{r_0}^2+\chi^2}}{\left({r_0}^2+\chi^2\right)^{2}} 
    \, , \quad 
    G_{\theta \theta} = \frac{r_0}{2 \sqrt{{r_0}^2+\chi^2} }
    \, , \quad 
    G_{\phi\phi} = \frac{r_0 \sin^2 \theta }{2 \sqrt{{r_0}^2+\chi^2} } 
    \, ,
\end{align}
and the above results leads to 
\begin{align}
\label{Norhops}
    \rho = 0 
    \, , \quad 
    p^\mathrm{radial} 
    = 
    - \frac{r_0}{\kappa^2 \left({r_0}^2+\chi^2\right)^{3/2}} 
    \, , \quad 
    p^\mathrm{angular} 
    = 
    \frac{r_0}{2 \kappa^2\left( {r_0}^2+\chi^2 \right)^{3/2} } 
    \, .
\end{align}
Because the energy density $\rho$ vanishes and $p^\mathrm{radial}$ is negative, all the energy conditions in Eq.~\eqref{Eq: energy cond2} are violated.
We note that the absolute value of $p^\mathrm{radial}$ reaches its maximum, $p^\mathrm{radial} = - 1/\kappa^2 r_0^2$, at $\chi = 0$.
The simplest case $N=0$ leads to violations of the energy conditions, and thus, it is necessary to study the non-trivial cases.
For a general choice of $N = N(\tau, \chi)$, the spacetime structure becomes complicated; however, as we will show, the corresponding energy conditions can be satisfied with an appropriate choice of $N(\tau, \chi)$.

\section{Future singularities in FLRW universe}
\label{SecV}

We further study the obtained wormhole solution by properly choosing $N(\tau,\chi)$. 
One possible direction for choosing a non-trivial form of $N$ is to investigate, in particular,
a wormhole embedded into the FLRW universe.
Furthermore, thanks to the four-scalar framework, we can not only embed a wormhole spacetime into a simple expanding universe but also consider an embedding into a universe with a finite future singularity.
For instance, it is interesting to study a spacetime in which there could be a singularity in our universe, but there is none in another universe. 
Even if there is a sudden future singularity, by travelling to another universe via the wormhole, we can avoid it and return to our universe afterwards. 

In this section, before exploring such a spacetime, we briefly review the finite future singularity appearance in the FLRW universe.
For the spatially flat FLRW universe with a scale factor $a(t)$, 
\begin{align}
\label{FLRW}
    ds^2 = - dt ^2 + a(t )^2 \sum_{i=1,2,3} \left( dx^i \right)^2 
    \, ,
\end{align}
the finite future singularities in cosmology are classified as follows \cite{Nojiri:2005sx}:
when $t \to t _0$, 
\begin{itemize}
    \item Type I (Big Rip) singularity: 
    $a(t )\rightarrow \infty$, $\rho\rightarrow\infty$ and $|p|\rightarrow\infty$.
    \item Type II (sudden) singularity: 
    $a(t ) \rightarrow \mathrm{const.}$ and $\rho \rightarrow \mathrm{const.}$, but $|p|\rightarrow\infty$.
    $a(t )$ and $\dot a(t )$ are finite, but $\ddot a(t )$ diverges. 
    \item Type III (big freeze) singularity: 
    $a(t ) \rightarrow \mathrm{const.}$, but $\rho\rightarrow\infty$ and $|p|\rightarrow\infty$. 
    $a(t )$ is finite, but $\dot a(t )$ diverges.
    \item Type IV (generalised sudden) singularity: 
    $a(t ) \rightarrow \mathrm{const.}$, $\rho \rightarrow \mathrm{const.}$, and $|p| \rightarrow \mathrm{const.}$, but some higher derivatives of the Hubble parameter $H$ diverge.
    $a(t )$, $\dot a(t )$ and $\ddot a(t )$ are finite, but higher derivatives of $a(t )$ diverge.
\end{itemize}
In addition to the above four types of singularities, the following has been proposed \cite{Dabrowski:2009kg}, 
\begin{itemize}
\item Type V ($w$) singularity: 
$w \rightarrow \infty$, but $p$ and $\rho$ are finite.
\end{itemize}
Here $w$ is the equation of state (EoS) parameter defined by $w=\frac{p}{\rho}$. 
In Type V singularity, the behaviour of the scale factor $a(t)$ is identical to that in Type II; that is, $a(t )$ and $\dot a(t )$ are finite, but $\ddot a(t )$ diverges. 
Type V singularity is just for the behaviour of the matter. 
Thus, Type I-IV singularities have completely classified the singular behaviours of spacetime.

We note that Type I singularity was first introduced in Ref.~\cite{Caldwell:2003vq}, which appears in the universe filled by the phantom fluid~\cite{Caldwell:1999ew}. 
Type II singularity was proposed in~\cite{Barrow:2004xh}. 
Type III and Type IV singularities were obtained by complementing the Type I and Type II singularities~\cite{Nojiri:2004pf, Nojiri:2005sx, Bouhmadi-Lopez:2006fwq}. 
As an example, let us assume that a singularity appears at $t =t _0$. 
When the Hubble parameter $H$ behaves as 
\begin{align}
\label{Halpha}
    H\sim H_1 \left( t _0 - t  \right)^\alpha
    \, ,
\end{align}
with constants $H_1>0$ and $\alpha$, Type I singularity appears for $\alpha\leq -1$, and Type III appears for $-1<\alpha<0$. 
Moreover, when $H$ behaves as 
\begin{align}
\label{HalphaB}
    H\sim H_0 + H_1 \left( t _0 - t  \right)^\alpha 
    \, ,
\end{align}
with constants $H_0>0$, $H_1$ and $\alpha$, Type II singularity appears for $0<\alpha<1$, and Type IV singularity appears for $\alpha>1$ where $\alpha$ is not an integer.

We often use the conformal time $\tau$ instead of $t$ in Eq.~\eqref{FLRW}. 
For the spatially flat FLRW universe, the conformal time $\tau$ is defined by, 
\begin{align}
\label{cFLRW}
    ds^2 = a(\tau)^2 \left(- d\tau^2 + \sum_{i=1,2,3} \left( dx^i \right)^2\right) 
    \, ,
\end{align}
For the wormhole spacetime as in Eq.~\eqref{TD1}, if $N$ does not depend on $\chi$ for large $\chi$, the spacetime is asymptotically identified with the spatially flat FLRW universe. 
Thus, we can identify $\tau$ in the wormhole spacetime~\eqref{TD1} with the conformal time $\tau$ in the FLRW spacetime~\eqref{cFLRW}.
With the above identifications, we can embed the wormhole geometry into the FLRW universe by imposing $N(\tau, \chi) \sim N(\tau)$ and $\e^{N(\tau)} \sim a(\tau)$.

\section{Embedding of wormhole in expanding universe}
\label{SecVI}

In this section, we formulate the embedding of a wormhole into an expanding universe by assuming $N$ does not depend on $\chi$ and only depends on the time $\tau$, $N=N(\tau)$. 
The cosmic time $t$ and conformal time $\tau$ are related as follows:
\begin{align}
\label{taut}
    dt = a d\tau = \e^{N}d\tau
    \, .
\end{align}
And as we discussed in the previous section, the embedded wormhole spacetime in Eq.~\eqref{TD1} is 
the asymptotically FLRW universe in Eq.~\eqref{FLRW}. 
In the FLRW universe~\eqref{FLRW}, the Hubble parameter in the cosmic time is given as 
\begin{align}
    H (t) = \frac{1}{a} \frac{da}{dt} = \frac{dN}{dt}
    \, .
\end{align}  
Note that $N$ is defined as a function of $\tau$ in Eq.~\eqref{TD1}.

For a universe with the finite future singularity, assuming Eq.~\eqref{Halpha}, we have
\begin{align}
\label{Nalpha}
    N\sim - \frac{H_1}{\alpha+1} \left( t_0 - t \right)^{\alpha+1}
    \, .
\end{align}
We find that $N$ is negative.
When $t\to t_0$, $N$ diverges in the case of Type I singularity ($\alpha\leq -1$) and vanishes in the case of Type III singularity ($-1<\alpha<0$). 
On the other hand, assuming Eq.~\eqref{HalphaB}, we find
When $H$ behaves as 
\begin{align}
\label{NalphaB}
    N\sim - \frac{H_1}{\alpha+1} \left( t_0 - t \right)^{\alpha+1} 
    + \mbox{regular terms} 
    \, .
\end{align}
In the cases of Type II ($0 < \alpha < 1$) and Type IV ($\alpha > 1$ and $\alpha$ is not an integer) singularities,
the second term, referred to as the regular term, dominates as $t \to t_0$, rather than the first term for the value of $N$ but the (higher-) derivative of the first term gives a singularity.

We note that in this model, the minimum of the throat radius is given by $\e^{N}r_0$ when $\chi=0$. 
If there are primordial wormholes before the inflation, the radius grows exponentially due to the expansion of the universe in the inflation epoch. 
Provided that the $e$-folding number of the inflation is $N=60$, the throat radius increases by $10^{26}$ times since $\e^{60}\sim 10^{26}$. 
The size of primordial wormholes may be constrained by the absence of any observational evidence for macroscopic wormholes in the present universe.

Next, we formulate the energy conditions by substituting $N=N(\tau)$ into Eq.~\eqref{NEinstein}.
We obtain the non-vanishing components of Einstein tensor  as follows:
\begin{align}
\label{NtEinstein}
    G_{\tau \tau} =&\, 3 \dot{N}^2 \, , \quad 
    G_{\tau \chi} = 0 \, , \quad 
    G_{\chi\chi} = -  \frac{2\left( r_0 + \sqrt{{r_0}^2+\chi^2}\right)}{\sqrt{{r_0}^2+\chi^2}} \ddot{N}
    - \frac{r_0 + \sqrt{{r_0}^2+\chi^2}}{\sqrt{{r_0}^2+\chi^2}} \dot{N}^2
    - \frac{{r_0}^2 + r_0 \sqrt{{r_0}^2+\chi^2}}{\left({r_0}^2+\chi^2\right)^{2}} \, , \nonumber \\
    G_{\theta \theta} =&\, -2 \left({r_0}^2+\chi^2\right) \ddot{N}
    - \left({r_0}^2+\chi^2\right) \dot{N}^2
    + \frac{r_0}{2 \sqrt{{r_0}^2+\chi^2} } \, , \quad 
    G_{\phi\phi} = \sin^2 \theta   G_{\theta \theta} \, .
\end{align}
Here $\dot N\equiv \frac{dN}{d\tau}$, etc. 
Then we find, 
\begin{align}
\label{Ntrhops}
    \rho 
    &=
    \frac{3\e^{-2 N(\tau)}\dot{N}^2}{\kappa^2} 
    \, , \quad 
    p^\mathrm{radial} 
    =
    \frac{\e^{-2 N}}{\kappa^2} \left\{ - 2 \ddot{N} - \dot{N}^2 - \frac{r_0}{\left({r_0}^2+\chi^2\right)^{3/2}} \right\} 
    \, , \nonumber \\ 
    p^\mathrm{angular} 
    &=
    \frac{\e^{-2 N}}{\kappa^2} \left\{ - 2 \ddot{N} - \dot{N}^2 + \frac{r_0}{2 \left( {r_0}^2+\chi^2 \right)^{3/2}} \right\} 
    \, .
\end{align}
Moreover, 
\begin{align}
\label{EC}
    \rho + p^\mathrm{radial} 
    &= 
    \frac{\e^{-2N}}{\kappa^2} \left\{ - 2\ddot N + 2 {\dot N}^2
    - \frac{r_0}{\left( {r_0}^2 + \chi^2\right)^\frac{3}{2}} \right\}
    \, , \nonumber \\
    \rho + p^\mathrm{angular} 
    &= 
    \frac{\e^{-2N}}{\kappa^2} \left\{ - 2\ddot N + 2 {\dot N}^2
    + \frac{r_0}{2\left( {r_0}^2 + \chi^2\right)^\frac{3}{2}} \right\}
    \, , \nonumber \\
    \rho +p^\mathrm{radial} + 2 p^\mathrm{angular} 
    &= 
    - \frac{6 \e^{-2N} \ddot N}{\kappa^2} 
    \, , \nonumber \\
    \rho - \left| p^\mathrm{radial} \right| 
    &=
    \frac{\e^{-2 N(\tau)}}{\kappa^2} \left[ 3\dot{N}^2 - \left| - 2 \ddot{N} - \dot{N}^2 - \frac{r_0}{\left({r_0}^2+\chi^2\right)^{3/2}} \right| \right] 
    \, , \nonumber \\
    \rho - \left| p^\mathrm{angular} \right| 
    &=
    \frac{\e^{-2 N(\tau)}}{\kappa^2} \left[ 3\dot{N}^2 - \left| - 2 \ddot{N} - \dot{N}^2 + \frac{r_0}{2 \left( {r_0}^2+\chi^2 \right)^{3/2}} \right| \right] 
    \, . 
\end{align}
Using the following relation between $N$ and $H$,
\begin{align}
\label{H}
    \dot{N}(\tau) \equiv \frac{dN}{d\tau} = \e^{N(t)} \frac{dN}{dt} = \e^{N(t)} H(t) 
    \, , \quad 
    \ddot{N}(\tau) = \e^{2N(t)} \left[ H^2(t) + \frac{dH(t)}{dt} \right] \, , 
\end{align}
we can rewrite Eq.~\eqref{EC} as follows:
\begin{align}
\label{EC2}
    \rho &= \frac{3H^2}{\kappa^2} > 0 \, , \quad 
    \rho + p^\mathrm{radial} 
    = 
    \frac{1}{\kappa^2} \left[ - \frac{dH}{dt} - \frac{r_0 \e^{-2N}}{\left( {r_0}^2 + \chi^2\right)^\frac{3}{2}} 
    \right] 
    \, , \nonumber \\
    \rho + p^\mathrm{angular} 
    &= 
    \frac{1}{\kappa^2} \left[ - \frac{dH}{dt} 
    + \frac{r_0 \e^{-2N}}{2\left( {r_0}^2 + \chi^2\right)^\frac{3}{2}} \right] 
    \, , \quad 
    \rho + p^\mathrm{radial} + 2 p^\mathrm{angular} 
    = 
    - \frac{6}{\kappa^2} \left[ H^2 + \frac{dH}{dt} \right] 
    \, , \nonumber \\
    \rho - \left| p^\mathrm{radial} \right| 
    &= 
    \frac{1}{\kappa^2} \left[ 3  H^2 - \left| 3 H^2(t) + 2 \frac{dH}{dt} + \frac{r_0 \e^{-2 N(\tau)}}{\left({r_0}^2+\chi^2\right)^{3/2}}  \right| \right] 
    \, , \nonumber \\
    \rho - \left| p^\mathrm{angular} \right| 
    &=
    \frac{1}{\kappa^2} \left[ 3 H^2 - \left| 3 H^2 + 2  \frac{dH}{dt}  - \frac{r_0 \e^{-2 N}}{2 \left( {r_0}^2+\chi^2 \right)^{3/2}} \right| \right] 
    \, ,
\end{align}

We note 
\begin{align}
    \frac{r_0}{\left( {r_0}^2 + \chi^2\right)^\frac{3}{2}} \leq \frac{1}{{r_0}^2}
    \, .
\end{align}
Thus, in the non-phantom universe, where $dH/dt<0$, NEC and WEC are satisfied if $- dH/dt> \frac{\e^{-2N}}{r_0^2}\geq \frac{\e^{-2N}r_0}{\left( {r_0}^2 + \chi^2\right)^\frac{3}{2}}$.
Because $\e^{-2N}\to 0$ in the late-time non-phantom universe, NEC and WEC are satisfied. 
In addition to the condition $-dH/dt > \e^{-2N}/r_0^2$, SEC is also satisfied if $H^2 + dH/dt<0$. 
The condition $H^2 + dH/dt < 0$ can be realised by a perfect fluid with the EoS parameter $w > -1/3$, which corresponds to a decelerating expansion.
Although the conditions required to satisfy DEC are somewhat complicated, it is, in principle, possible to fulfil them.
We conclude that in a universe exhibiting a future singularity, the energy conditions are generally violated, as in the standard FLRW case.

We find that all the energy conditions can be satisfied if $-dH/dt > \e^{-2N}/{r_0}^2$ and $H^2 + dH/dt < 0$,
which can occur when the physical radius of the wormhole throat, $r_0 \e^{N}$, is comparable to or larger than the Hubble radius.
That is,
\begin{align}
\label{Eq: wh term}
    \frac{r_0 \e^{-2N}}{\left( {r_0}^2 + \chi^2\right)^\frac{3}{2}} 
    = 
    \frac{1}{ \left( r_0\e^{N}\right)^2} \frac{1}{\left[ 1 + \left( \frac{\chi}{r_0} \right)^2\right]^\frac{3}{2}} 
    \, ,
\end{align}
and $({r_0} \e^{N})^{-2} <-dH/dt \sim H^2$ for the case the standard power-law expansion case, $H\propto 1/t$, which is realised when the EoS parameter $w$ is constant and greater than $-1$.
In this case, the observable universe would be contained within the wormhole.
Such a configuration could naturally arise in the early universe, suggesting that the topology of the primordial universe might have been non-trivial.
At later time, as the universe expands ($N \to \infty$), the throat radius $r_0 \e^{N}$ increases continuously,
implying that the entire universe is eventually engulfed by the wormhole.

We emphasise that, in any case, the energy conditions can be satisfied in our model. 
As we mentioned, even if any of the conditions is violated, there is no instability related to the scalar fields in our four-scalar model~\eqref{Eq. acg1} with constraints~\eqref{Eq: cnstrnt1}. 
For the choice of $N=N(\tau)$ as in Eq.~\eqref{taut}, the non-vanishing components of the Ricci tensor in Eq.~\eqref{TD3} are reduced into the following forms: 
\begin{align}
\label{TD3B}
    R_{\tau\tau} 
    &= 
    -3 \ddot{N} 
    \, , \nonumber \\ 
    R_{\chi\chi} 
    &= 
    \frac{r_0 +\sqrt{{r_0}^2+\chi^2}}{\sqrt{{r_0}^2+\chi^2}} \ddot{N} 
    -3 N^{(0,2)} 
    + \frac{2\left(r_0 +\sqrt{{r_0}^2+\chi^2}\right)}{\sqrt{{r_0}^2+\chi^2}} {\dot{N}}^2
    - \frac{{r_0}^2+r_0 \sqrt{{r_0}^2+\chi^2}}{\left({r_0}^2+\chi^2\right)^{2}} 
    \, , \nonumber \\
    R_{\theta \theta} 
    &=
    \frac{1}{\sin^2 \theta }   R_{\phi\phi} 
    \nonumber \\
    &= 
    \left({r_0}^2+\chi^2\right) \ddot{N} 
    + 2 \left({r_0}^2+\chi^2\right) {\dot{N}}^2 
    + \frac{1}{2} \frac{r_0}{\sqrt{{r_0}^2+\chi^2}}  
    \, ,
\end{align}
and other components vanish.
Replacing the coordinates $(\tau,\chi,\theta,\phi)$ with four scalar fields $\left(\phi_0, \phi_1, \phi_2, \phi_3 \right)$ in Eq.~\eqref{TD3B} and using Eq.~\eqref{Eqs5B}, we obtain a non-linear $\sigma$ model~\eqref{Eq. acg2} that realises the dynamical wormhole in Eq.~\eqref{TD1} with $N=N(\tau)$.

\section{Energy conditions for constant EoS parameter}
\label{SecVII}

In this section, for the wormhole spacetime embedded in the expanding universe, we show graphically how the energy conditions can be satisfied. 
In the following, we assume a fluid with the constant EoS parameter $w$ and consider the following three cases:
\begin{enumerate}
\item\label{i1}
The case $-1<w<-1/3$ corresponds to the universe filled with the quintessence, and there is no finite future singularity for $-1<w$.
Including the boundary case of $w=-1/3$, we consider the range of $-1<w\leq-1/3$.
The Hubble parameter in the cosmic time $H(t)$ is given by 
\begin{align}
\label{QntcH}
    H(t) 
    = \frac{dN}{dt} 
    = \frac{2}{3(1+w) t} = H_0 \cdot \frac{t_0}{t} 
    \, , \quad
    H_{0} = \frac{2}{3(1+w)t_0}
    \, .
\end{align}
Here, $t_0$ is a constant. 
And we obtain $dHH/dt$ and $N$,
\begin{align}
    \frac{dH}{dt} 
    =
    - \frac{3(1+w)}{2} \cdot H_0^2 \cdot \left(\frac{t}{t_0}\right)^{-2}
    \, , \quad
    N(t) 
    =
    \ln \left( \frac{t}{t_0}  \right)^{\frac{2}{3(1+w)}}
    \, .
\end{align}
Solving Eq.~\eqref{taut} for the obtained $N(t)$, we can rewrite $t$ by $\tau$, 
\begin{align}
    \frac{t}{t_0}  
    &= 
    \left[ \frac{H_{0}(1+3w)}{2} \left(\tau - \tau_0 \right) \right]^{\frac{3(1+w)}{1+3w}}
    \, , \quad
    N(\tau)
    =
    \ln \left[  \frac{H_{0}(1+3w)}{2} \left(\tau - \tau_0 \right) \right]^{\frac{2}{1+3w}}
    \, .
\end{align}

\item\label{i2} 
The case $w=-1$ corresponds to the universe with the cosmological constant.
The Hubble parameter is constant
\begin{align}
    H(t)=H_0 
    \quad 
    \left( H_0:\ \mbox{constant}\right)
    \, ,
\end{align}
and thus,
\begin{align}
    \frac{dH}{dt} = 0
    \, , \quad
    N(t) =  H_0 (t-t_0)
    \, .
\end{align}
Using Eq.~\eqref{taut}, we find
\begin{align}
    t-t_0
    &= 
    - \frac{1}{H_0} \ln[ - H_0(\tau - \tau_0)]
    \, , \quad
    N(\tau)
    =
    -\ln[ - H_0(\tau - \tau_0)]
    \, .
\end{align}

\item\label{i3} 
The case $w<-1$ corresponds to the universe filled with the phantom dark energy.
The Hubble parameter is given by
\begin{align}
\label{HbbleBR}
    H(t) 
    = \frac{2}{3(1+w) (t - t_s)} 
    = H_0 \cdot \left( \frac{t - t_s}{t_0 - t_s} \right)^{-1} 
    \, , \quad
    H_0 \equiv \frac{2}{3 (1+w)\left(t_0 - t_s\right)}
    \, .
\end{align}
Here, $t_s$ is the time when the (Big Rip) singularity shows up.
And, we obtain
\begin{align}
    \frac{dH}{dt} 
    =
    - \frac{3(1+w)}{2} \cdot H_0^2 \cdot
    \left( \frac{t - t_s}{t_0 - t_s} \right)^{-2}
    \, , \quad
    N(t)
    =
    \ln \left( \frac{t - t_s}{t_0 - t_s} \right)^{\frac{2}{3(1+w)}}
\end{align}
Using Eq.~\eqref{taut}, we find
\begin{align}
\begin{split}
    \frac{t - t_s}{t_0 - t_s} 
    &= 
    \left[ \frac{H_{0}(1+3w)}{2} (\tau - \tau_0)\right]^{\frac{3(1+w)}{1+3w}} 
    \, , \quad
    N(\tau)
    = 
    \ln \left[ \frac{H_{0}(1+3w)}{2} (\tau - \tau_0)\right]^{\frac{2}{1+3w}}
    \, .
\end{split}
\end{align}
In the later analysis, we set $t_0 = 0$ for convenience.
\end{enumerate}

\subsection{Case 1: $-1<w\leq -1/3$}

In the case $-1<w\leq -1/3$, Eq.~\eqref{EC2} yields
\begin{align}
\label{EC case1}
    \rho =& 
    \frac{3 {H_0}^2}{\kappa^2} \left( \frac{t}{t_0} \right)^{-2} 
    \, , \nonumber \\ 
    \rho + p^\mathrm{radial} =& \frac{{H_0}^2 }{\kappa^2} \left[ \frac{3(1+w)}{2}   \left(\frac{t}{t_0}\right)^{-2} 
 - \frac{1}{(r_0 H_0)^2} 
    \frac{1}{\left[1+ \left(\frac{\chi}{r_0}\right)^2\right]^\frac{3}{2}} \left( \frac{t}{t_0} \right)^{-\frac{4}{3(1+w)}} 
    \right] 
    \, , \nonumber \\
    \rho + p^\mathrm{angular} 
    =& 
    \frac{{H_0}^2}{\kappa^2} \left[ \frac{3(1+w)}{2}    \left(\frac{t}{t_0}\right)^{-2}
    + \frac{1}{2} \frac{1}{(r_0 H_0)^2} 
    \frac{1}{\left[1+ \left(\frac{\chi}{r_0}\right)^2\right]^\frac{3}{2}} \left( \frac{t}{t_0} \right)^{-\frac{4}{3(1+w)}} 
    \right] 
    \, , \nonumber \\
    \rho + p^\mathrm{radial} + 2 p^\mathrm{angular} 
    =& 
    \frac{3 (1+3w) {H_0}^2}{\kappa^2} \left(\frac{t}{t_0}\right)^{-2} 
    \, , \nonumber \\
    \rho - | p^\mathrm{radial}| 
    =& 
    \frac{{H_0}^2}{\kappa^2} \left[ 3 \left( \frac{t}{t_0}\right)^{-2} - \left| 3w \left(\frac{t}{t_0}\right)^{-2} - \frac{1}{(r_0 H_0)^2} 
    \frac{1}{\left[1+ \left(\frac{\chi}{r_0}\right)^2\right]^\frac{3}{2}} \left( \frac{t}{t_0} \right)^{-\frac{4}{3(1+w)}} \right| \right] 
    \, , \nonumber \\
    \rho -| p^\mathrm{angular}| 
    =& 
    \frac{{H_0}^2}{\kappa^2} \left[ 3 \left( \frac{t}{t_0}\right)^{-2} - \left| 3w \left(\frac{t}{t_0}\right)^{-2} + \frac{1}{2} \frac{1}{(r_0 H_0)^2} 
    \frac{1}{\left[1+ \left(\frac{\chi}{r_0}\right)^2\right]^\frac{3}{2}} \left( \frac{t}{t_0} \right)^{-\frac{4}{3(1+w)}} \right| \right] 
    \, .
\end{align}
Note that the factor $r_0 H_0$ represents the ratio of the wormhole throat scale $r_0$ to the Hubble radius $1/H_0$.
For parameters $w, r_0 H_0$, we plot the above six quantities as functions of $(t/t_0, \chi/r_0)$.

\begin{figure}[htbp]
\centering
    \begin{subfigure}[b]{0.3\textwidth}
    \centering
    \includegraphics[width=\textwidth]{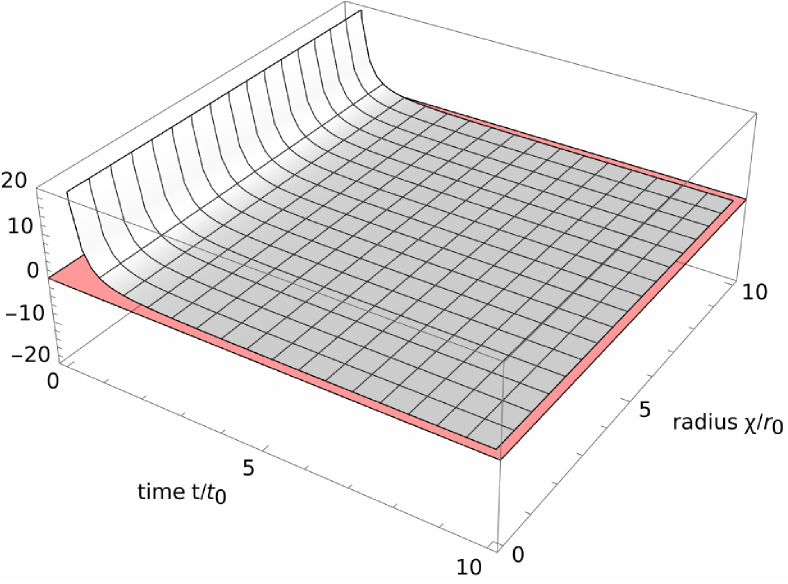}
    \caption{Plot of $\rho$ normalised by $\frac{{H_0}^2}{\kappa^2}$ with $r_0 H_0 = 10^{-2}$, $w=-1/3$}
    \label{fig1:sub1}
\end{subfigure}
\hfill
    \begin{subfigure}[b]{0.3\textwidth}
    \centering
    \includegraphics[width=\textwidth]{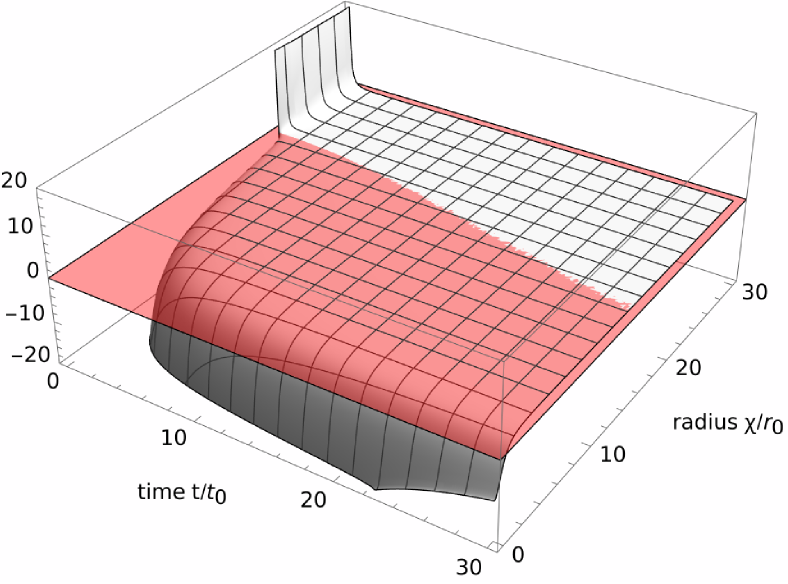}
    \caption{Plot of $\rho + p^\mathrm{radial} $ normalised by $\frac{{H_0}^2}{\kappa^2}$ with $r_0 H_0 = 10^{-2}$, $w=-1/3$}
    \label{fig1:sub2}
\end{subfigure}
\hfill
    \begin{subfigure}[b]{0.3\textwidth}
    \centering
    \includegraphics[width=\textwidth]{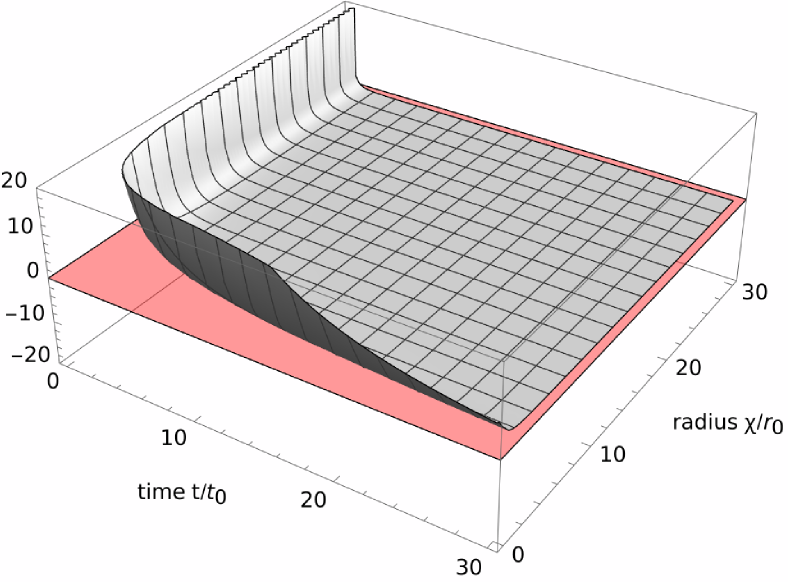}
    \caption{Plot of $\rho+p^\mathrm{angular} $ normalised by $\frac{{H_0}^2}{\kappa^2}$ with $r_0 H_0 = 10^{-2}$, $w=-1/3$}
    \label{fig1:sub3}
\end{subfigure}
\hfill
\begin{subfigure}[b]{0.3\textwidth}
    \centering
    \includegraphics[width=\textwidth]{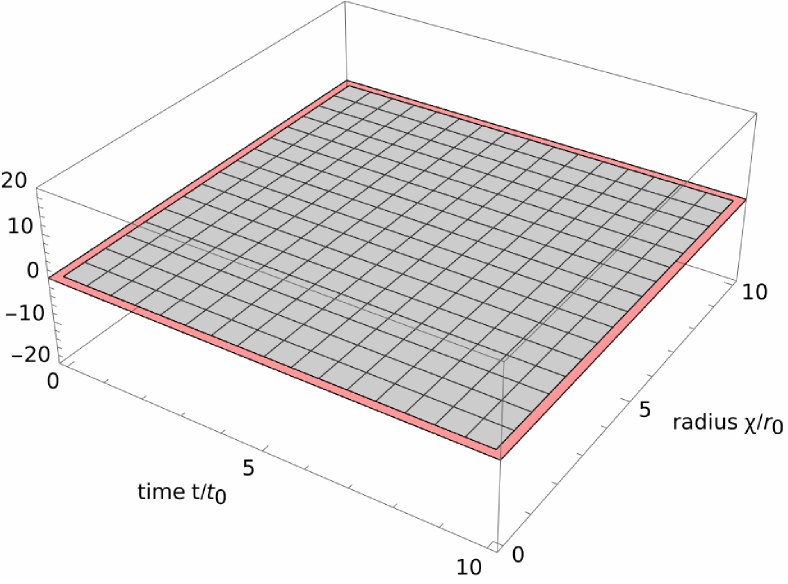}
    \caption{Plot of $\rho + p^\mathrm{radial} + 2 p^\mathrm{angular} $ normalised by $\frac{{H_0}^2}{\kappa^2}$ with $r_0 H_0 = 10^{-2}$, $w=-1/3$}
    \label{fig1:sub4}
\end{subfigure}
\hfill
\begin{subfigure}[b]{0.3\textwidth}
    \centering
    \includegraphics[width=\textwidth]{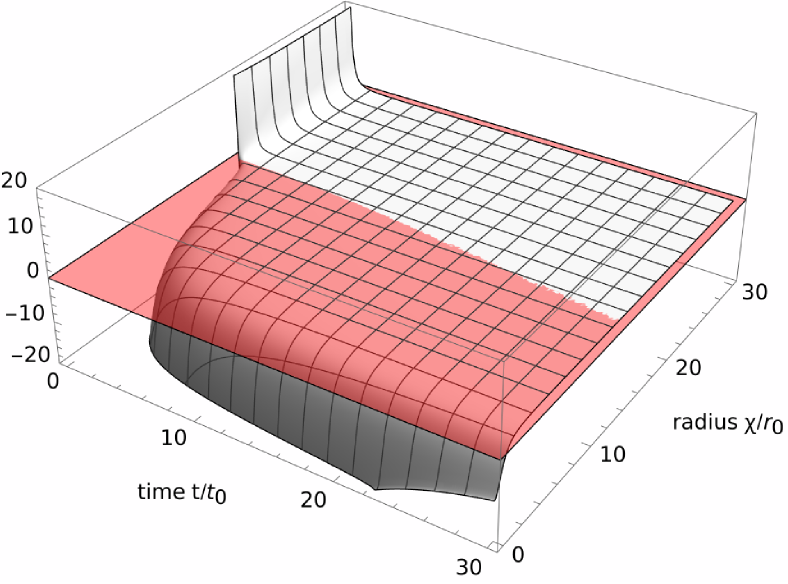}
    \caption{Plot of $\rho - \left| p^\mathrm{radial} \right|$ normalised by $\frac{{H_0}^2}{\kappa^2}$ with $r_0 H_0 = 10^{-2}$, $w=-1/3$}
    \label{fig1:sub5}
\end{subfigure}
\hspace{20pt}
\begin{subfigure}[b]{0.3\textwidth}
    \centering
    \includegraphics[width=\textwidth]{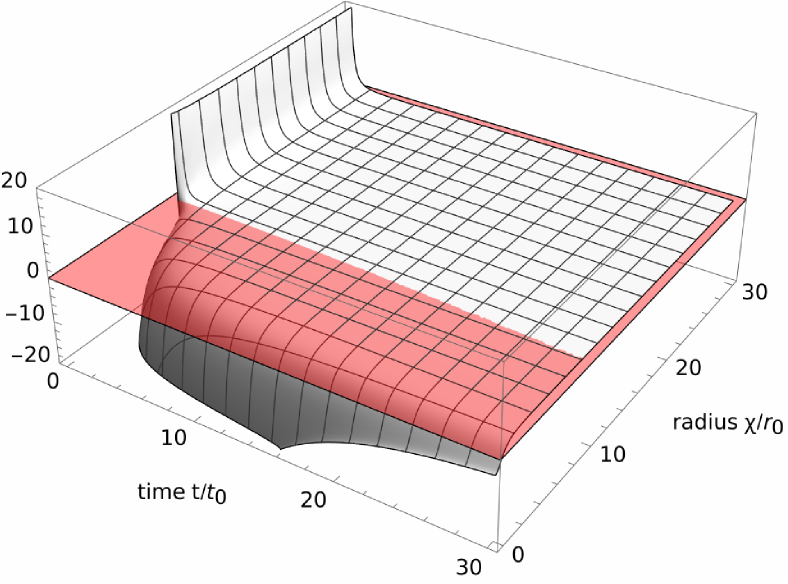}
    \caption{Plot of $\rho - \left| p^\mathrm{angular} \right|$ normalised by $\frac{{H_0}^2}{\kappa^2}$ with $r_0 H_0 = 10^{-2}$, $w=-1/3$}
    \label{fig1:sub6}
\end{subfigure}
\caption{The behaviours of $\rho$ \eqref{fig1:sub1}, $\rho + p^\mathrm{radial} $ \eqref{fig1:sub2}, $\rho + p^\mathrm{angular}$ \eqref{fig1:sub3}, 
$\rho + p^\mathrm{radial} + 2 p^\mathrm{angular} $ \eqref{fig1:sub4}, 
$\rho - \left| p^\mathrm{radial}\right|$ \eqref{fig1:sub5} and $\rho - \left| p^\mathrm{angular} \right|$ \eqref{fig1:sub6}, 
when we use the normalisation of $\frac{{H_0}^2}{\kappa^2}$ 
with $r_0 H_0 = 10^{-2}$, $w=-1/3$. 
The red plane indicates $z=0$, and each quantity takes positive values above this plane.
All the energy conditions are violated, but they all might be satisfied in the late time.}
\label{fig1:overall}
\end{figure}

\begin{figure}[htbp]
\centering
\begin{subfigure}[b]{0.3\textwidth}
    \centering
    \includegraphics[width=\textwidth]{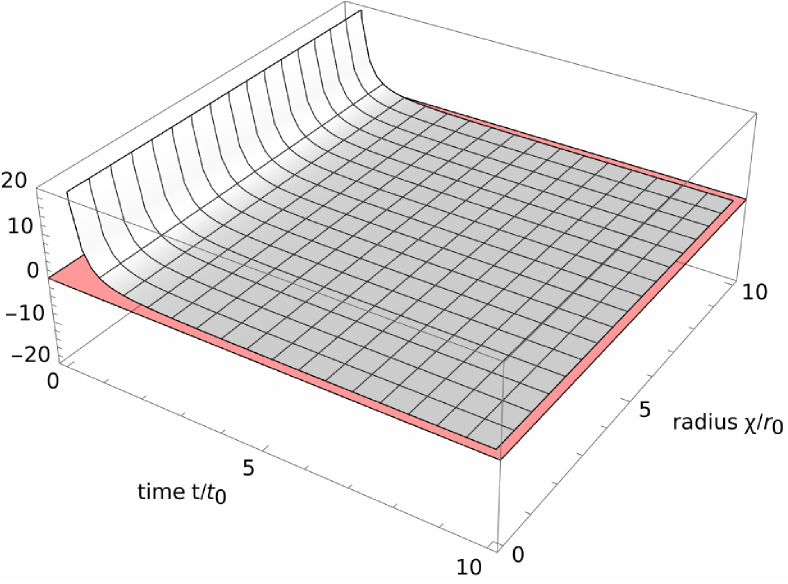}
    \caption{Plot of $\rho$ normalised by $\frac{{H_0}^2}{\kappa^2}$ with $r_0 H_0 = 10^{0}$, $w=-1/3$}
    \label{fig3:sub1}
\end{subfigure}
\hfill
\begin{subfigure}[b]{0.3\textwidth}
    \centering
    \includegraphics[width=\textwidth]{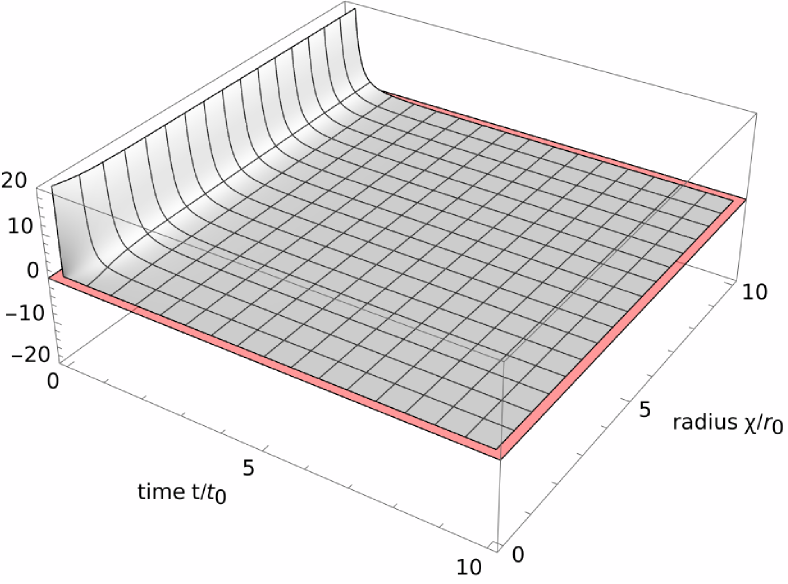}
    \caption{Plot of $\rho + p^\mathrm{radial} $ normalised by $\frac{{H_0}^2}{\kappa^2}$ with $r_0 H_0 = 10^{0}$, $w=-1/3$}
    \label{fig3:sub2}
\end{subfigure}
\hfill
\begin{subfigure}[b]{0.3\textwidth}
    \centering
    \includegraphics[width=\textwidth]{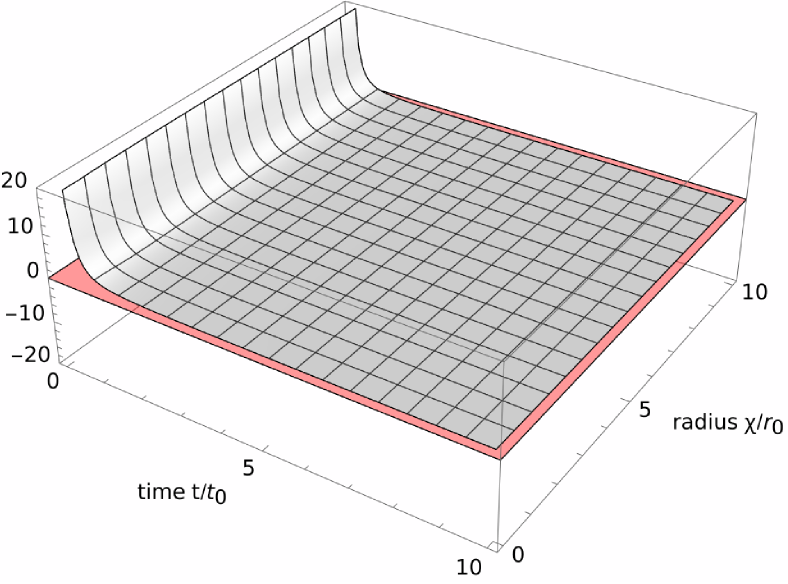}
    \caption{Plot of $\rho+p^\mathrm{angular} $ normalised by $\frac{{H_0}^2}{\kappa^2}$ with $r_0 H_0 = 10^{0}$, $w=-1/3$}
    \label{fig3:sub3}
\end{subfigure}
\hfill
\begin{subfigure}[b]{0.3\textwidth}
    \centering
    \includegraphics[width=\textwidth]{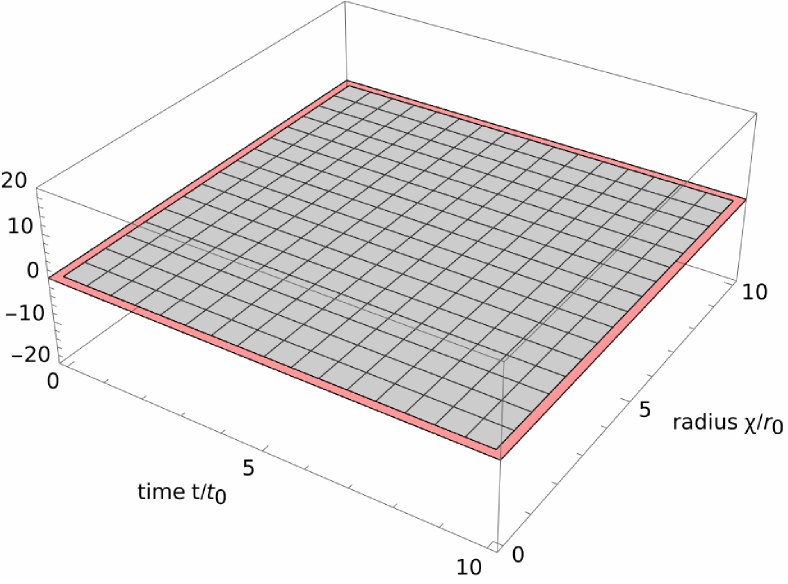}
    \caption{Plot of $\rho + p^\mathrm{radial} + 2 p^\mathrm{angular} $ normalised by $\frac{{H_0}^2}{\kappa^2}$ with $r_0 H_0 = 10^{0}$, $w=-1/3$}
    \label{fig3:sub4}
\end{subfigure}
\hfill
\begin{subfigure}[b]{0.3\textwidth}
    \centering
    \includegraphics[width=\textwidth]{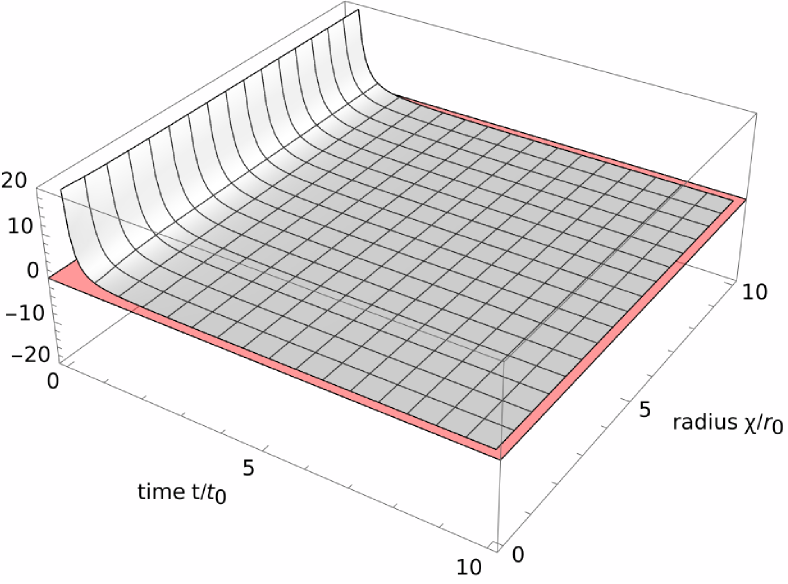}
    \caption{Plot of $\rho - \left| p^\mathrm{radial} \right|$ normalised by $\frac{{H_0}^2}{\kappa^2}$ with $r_0 H_0 = 10^{0}$, $w=-1/3$}
    \label{fig3:sub5}
\end{subfigure}
\hfill
\begin{subfigure}[b]{0.3\textwidth}
    \centering
    \includegraphics[width=\textwidth]{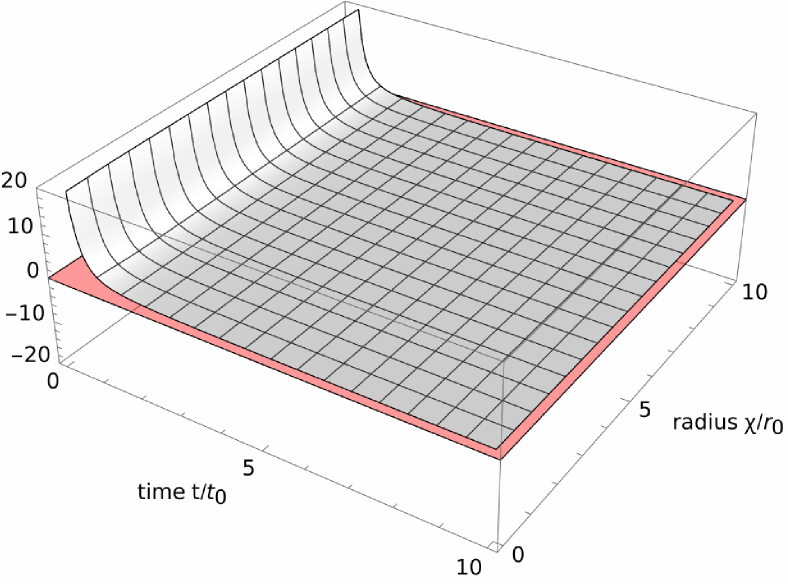}
    \caption{Plot of $\rho - \left| p^\mathrm{angular} \right|$ normalised by $\frac{{H_0}^2}{\kappa^2}$ with $r_0 H_0 = 10^{0}$, $w=-1/3$}
    \label{fig3:sub6}
\end{subfigure}
\caption{The behaviours of $\rho$ \eqref{fig3:sub1}, $\rho + p^\mathrm{radial} $ \eqref{fig3:sub2}, $\rho + p^\mathrm{angular}$ \eqref{fig3:sub3}, 
$\rho + p^\mathrm{radial} + 2 p^\mathrm{angular} $ \eqref{fig3:sub4}, 
$\rho - \left| p^\mathrm{radial}\right|$ \eqref{fig3:sub5} and $\rho - \left| p^\mathrm{angular} \right|$ \eqref{fig3:sub6}, 
when we use the normalisation of $\frac{{H_0}^2}{\kappa^2}$ 
with $r_0 H_0 = 10^{0}$, $w=-1/3$. 
The red plane indicates $z=0$, and each quantity takes positive values above this plane.
All the energy conditions do not seem to be violated.}
\label{fig3:overall}
\end{figure}

Fig.~\ref{fig1:overall} shows $\rho$, $\rho + p^\mathrm{radial} $, $\rho + p^\mathrm{radial} + 2 p^\mathrm{angular} $, $\rho - \left| p^\mathrm{radial}\right|$ and $\rho - \left| p^\mathrm{angular} \right|$  for $r_0 H_0 = 10^{-2}$, $w=-1/3$.
We normalised all quantities by the factor ${H_0}^2/\kappa^2$.
All the energy conditions are violated, but they all might be satisfied in the late time.
This could always be true when $w\geq - 1/3$. 
Fig.~\ref{fig3:overall} shows the same six quantities for $r_0 H_0 = 10^{0}$ and $w=-1/3$. 
Comparing Fig.~\ref{fig3:overall} ($r_0 H_0 = 10^{0}$) with Fig.~\ref{fig1:overall} ($r_0 H_0 = 10^{-2}$), we find the impact of the $r_0$.
All the energy conditions do not appear to be violated. 
When $w\geq - 1/3$, this result could be because the radius of the wormhole throat could be larger than or equal to the Hubble radius. 

\begin{figure}[htbp]
\centering
\begin{subfigure}[b]{0.3\textwidth}
    \centering
    \includegraphics[width=\textwidth]{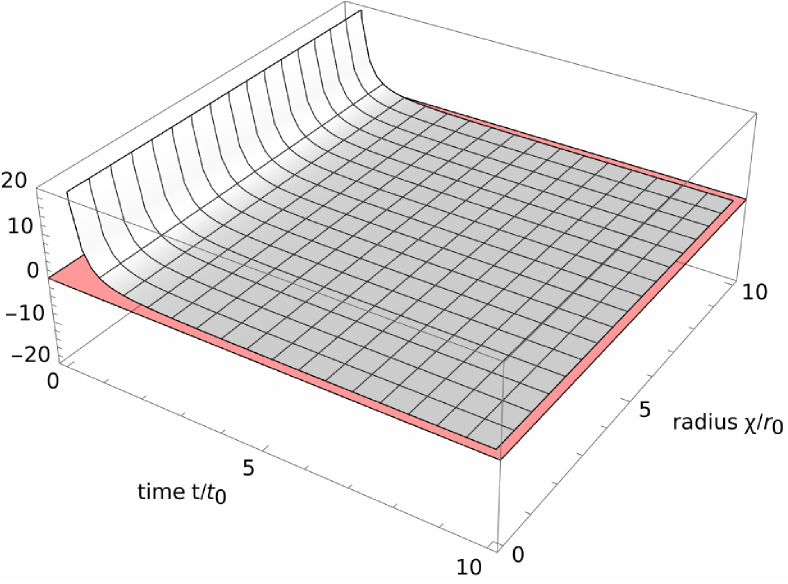}
    \caption{Plot of $\rho$ normalised by $\frac{{H_0}^2}{\kappa^2}$ with $r_0 H_0 = 10^{-2}$, $w=-2/3$}
    \label{fig2:sub1}
\end{subfigure}
\hfill
    \begin{subfigure}[b]{0.3\textwidth}
    \centering
    \includegraphics[width=\textwidth]{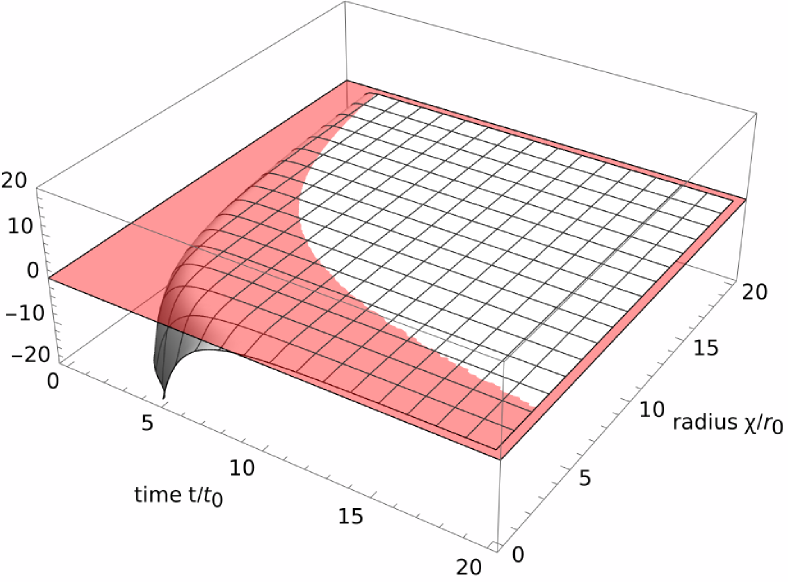}
    \caption{Plot of $\rho + p^\mathrm{radial} $ normalised by $\frac{{H_0}^2}{\kappa^2}$ with $r_0 H_0 = 10^{-2}$, $w=-2/3$}
    \label{fig2:sub2}
\end{subfigure}
\hfill
\begin{subfigure}[b]{0.3\textwidth}
    \centering
    \includegraphics[width=\textwidth]{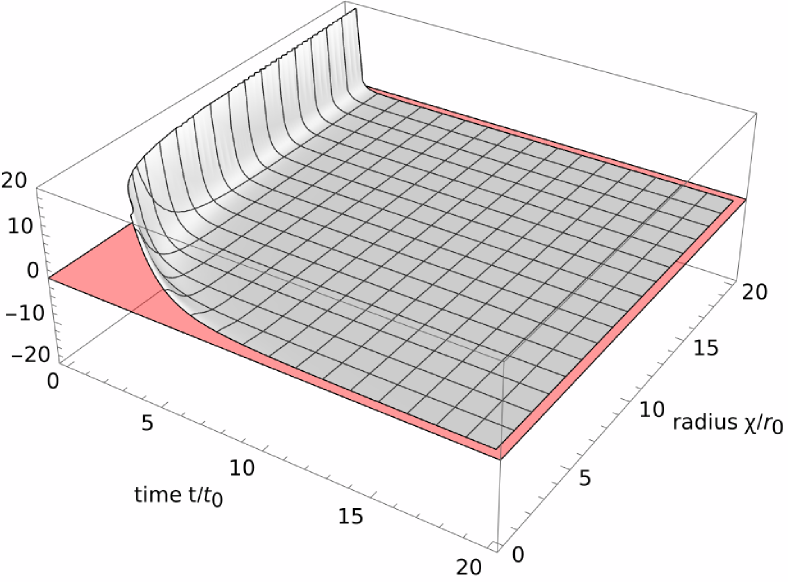}
    \caption{Plot of $\rho+p^\mathrm{angular} $ normalised by $\frac{{H_0}^2}{\kappa^2}$ with $r_0 H_0 = 10^{-2}$, $w=-2/3$}
    \label{fig2:sub3}
\end{subfigure}
\hfill
\begin{subfigure}[b]{0.3\textwidth}
    \centering
    \includegraphics[width=\textwidth]{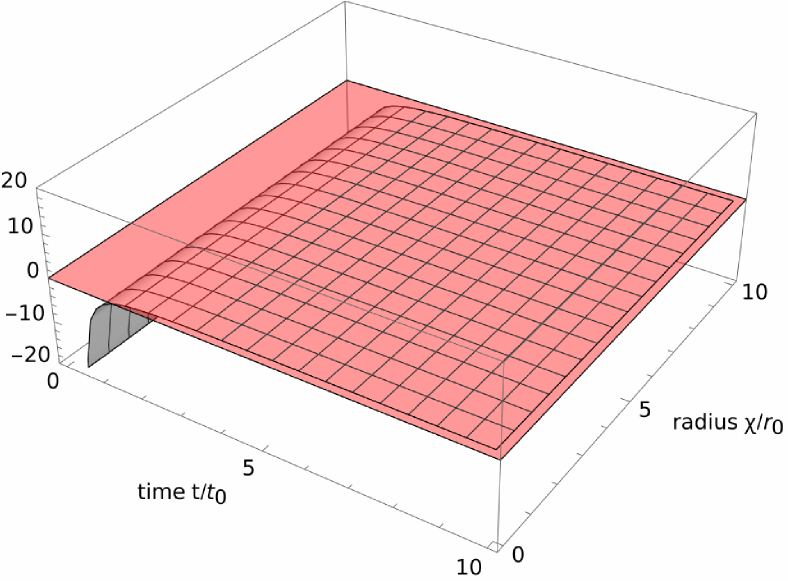}
    \caption{Plot of $\rho + p^\mathrm{radial} + 2 p^\mathrm{angular} $ normalised by $\frac{{H_0}^2}{\kappa^2}$ with $r_0 H_0 = 10^{-2}$, $w=-2/3$}
    \label{fig2:sub4}
\end{subfigure}
\hfill
\begin{subfigure}[b]{0.3\textwidth}
    \centering
    \includegraphics[width=\textwidth]{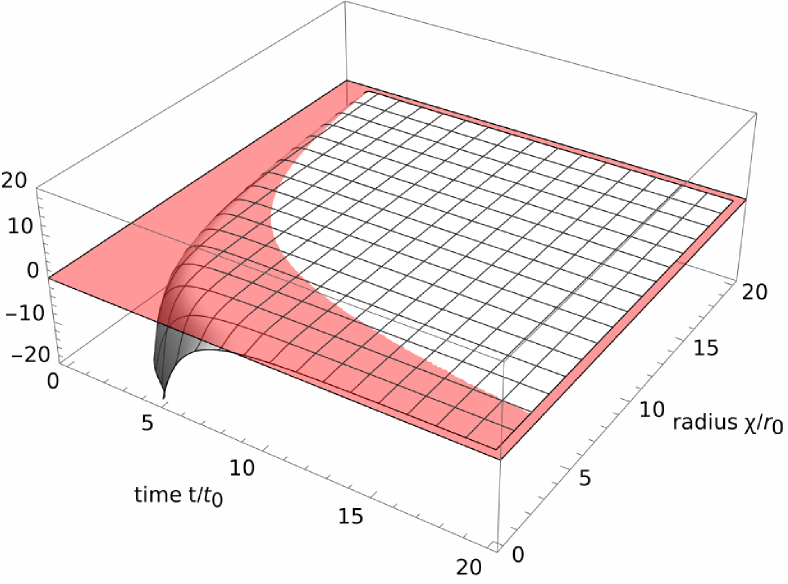}
    \caption{Plot of $\rho - \left|p^\mathrm{radial} \right|$ normalised by $\frac{{H_0}^2}{\kappa^2}$ with $r_0 H_0 = 10^{-2}$, $w=-2/3$}
    \label{fig2:sub5}
\end{subfigure}
\hfill
\begin{subfigure}[b]{0.3\textwidth}
    \centering
    \includegraphics[width=\textwidth]{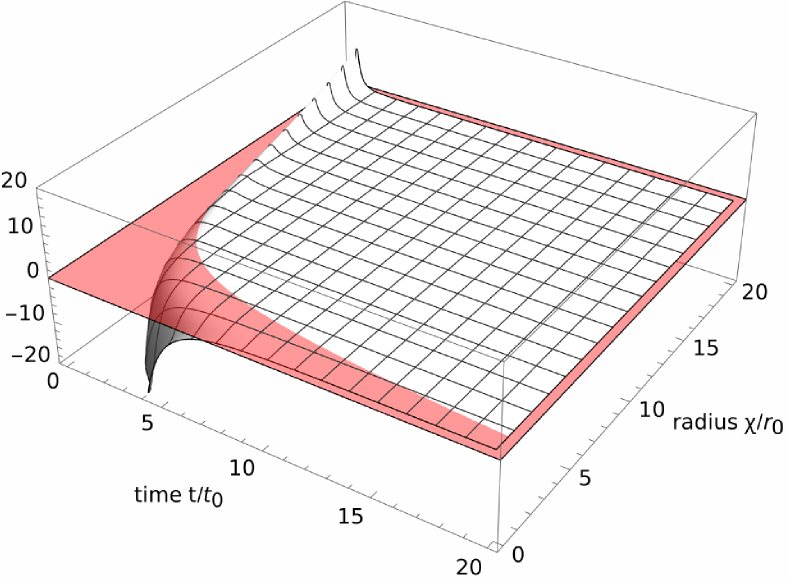}
    \caption{Plot of $\rho - \left| p^\mathrm{angular} \right|$ normalised by $\frac{{H_0}^2}{\kappa^2}$ with $r_0 H_0 = 10^{-2}$, $w=-2/3$}
    \label{fig2:sub6}
\end{subfigure}
\caption{The behaviours of $\rho$ \eqref{fig2:sub1}, $\rho + p^\mathrm{radial} $ \eqref{fig2:sub2}, $\rho + p^\mathrm{angular}$ \eqref{fig2:sub3}, 
$\rho + p^\mathrm{radial} + 2 p^\mathrm{angular} $ \eqref{fig2:sub4}, 
$\rho - \left| p^\mathrm{radial}\right|$ \eqref{fig2:sub5} and $\rho - \left| p^\mathrm{angular} \right|$ \eqref{fig2:sub6}, 
when we use the normalisation of $\frac{{H_0}^2}{\kappa^2}$ 
with $r_0 H_0 = 10^{-2}$, $w=-2/3$. 
All the energy conditions are violated, but they all might be recovered in the late time, except SEC, which is always violated.}
\label{fig2:overall}
\end{figure}

\begin{figure}[htbp]
\centering
\begin{subfigure}[b]{0.3\textwidth}
    \centering
    \includegraphics[width=\textwidth]{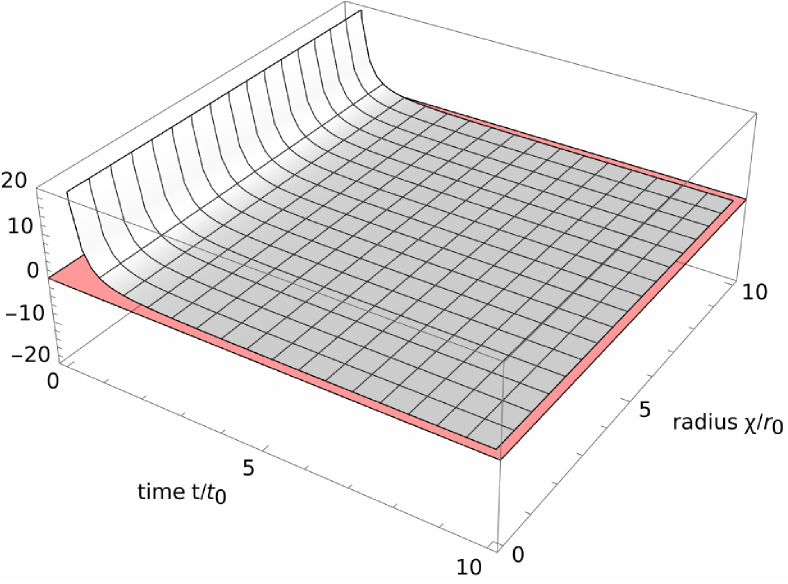}
    \caption{Plot of $\rho$ normalised by $\frac{{H_0}^2}{\kappa^2}$ with $r_0 H_0 = 10^{0}$, $w=-2/3$}
    \label{fig4:sub1}
\end{subfigure}
\hfill
\begin{subfigure}[b]{0.3\textwidth}
    \centering
    \includegraphics[width=\textwidth]{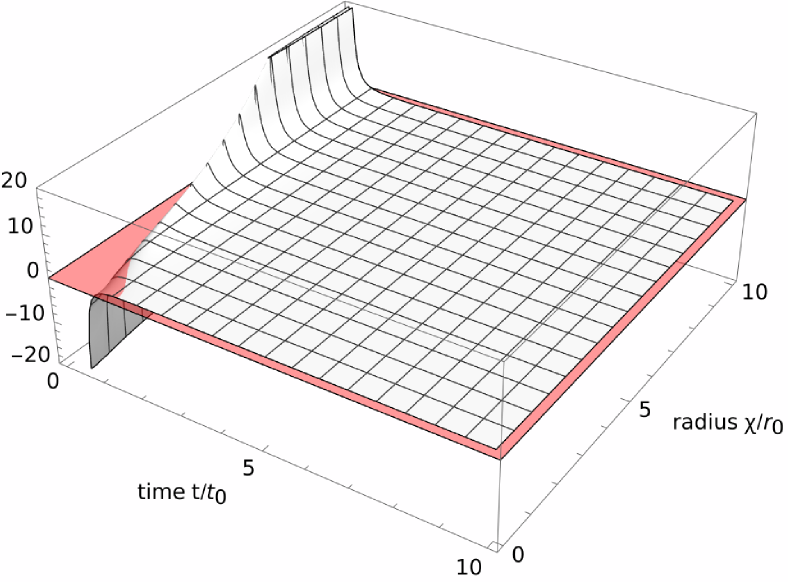}
    \caption{Plot of $\rho + p^\mathrm{radial} $ normalised by $\frac{{H_0}^2}{\kappa^2}$ with $r_0 H_0 = 10^{0}$, $w=-2/3$}
    \label{fig4:sub2}
\end{subfigure}
\hfill
\begin{subfigure}[b]{0.3\textwidth}
    \centering
    \includegraphics[width=\textwidth]{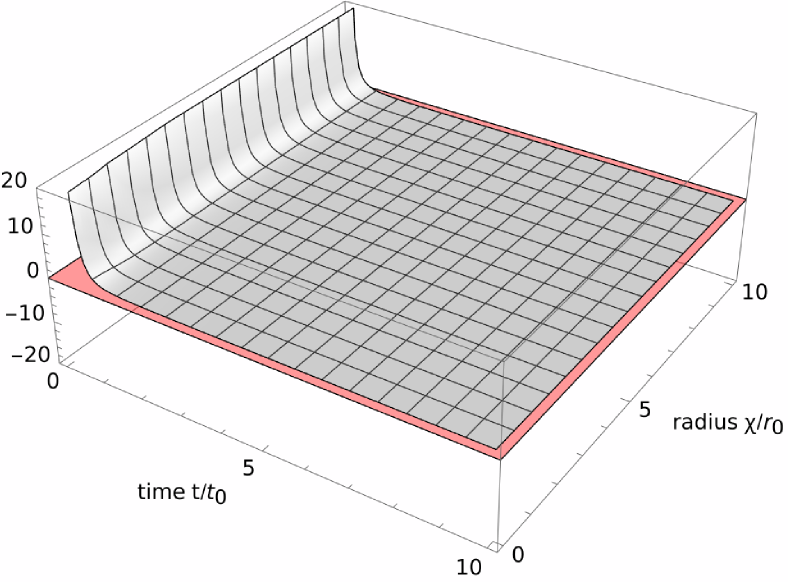}
    \caption{Plot of $\rho+p^\mathrm{angular} $ normalised by $\frac{{H_0}^2}{\kappa^2}$ with $r_0 H_0 = 10^{0}$, $w=-2/3$}
    \label{fig4:sub3}
\end{subfigure}
\hfill
\begin{subfigure}[b]{0.3\textwidth}
    \centering
    \includegraphics[width=\textwidth]{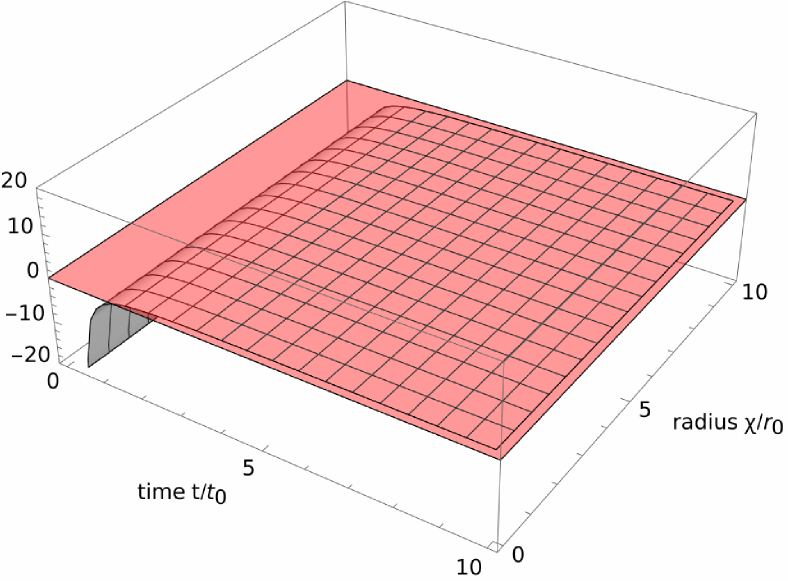}
    \caption{Plot of $\rho + p^\mathrm{radial} + 2 p^\mathrm{angular} $ normalised by $\frac{{H_0}^2}{\kappa^2}$ with $r_0 H_0 = 10^{0}$, $w=-2/3$}
    \label{fig4:sub4}
\end{subfigure}
\hfill
\begin{subfigure}[b]{0.3\textwidth}
    \centering
    \includegraphics[width=\textwidth]{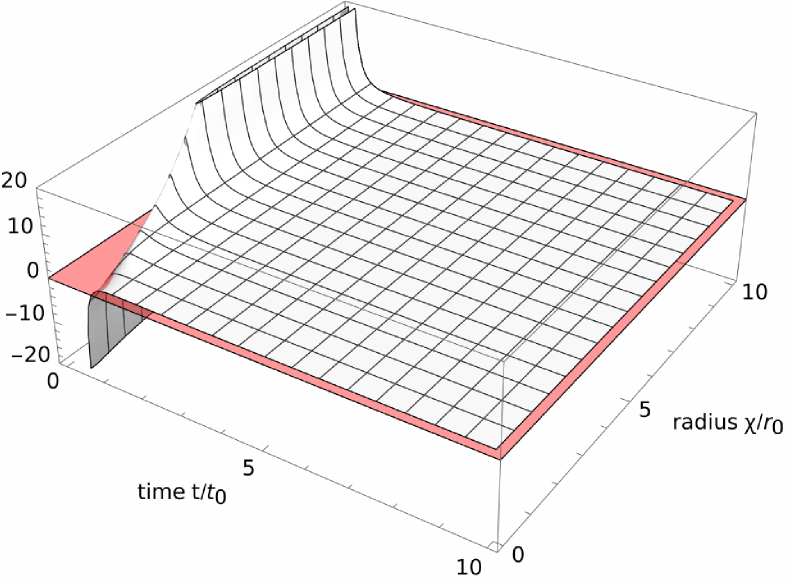}
    \caption{Plot of $\rho - \left| p^\mathrm{radial} \right|$ normalised by $\frac{{H_0}^2}{\kappa^2}$ with $r_0 H_0 = 10^{0}$, $w=-2/3$}
    \label{fig4:sub5}
\end{subfigure}
\hfill
\begin{subfigure}[b]{0.3\textwidth}
    \centering
    \includegraphics[width=\textwidth]{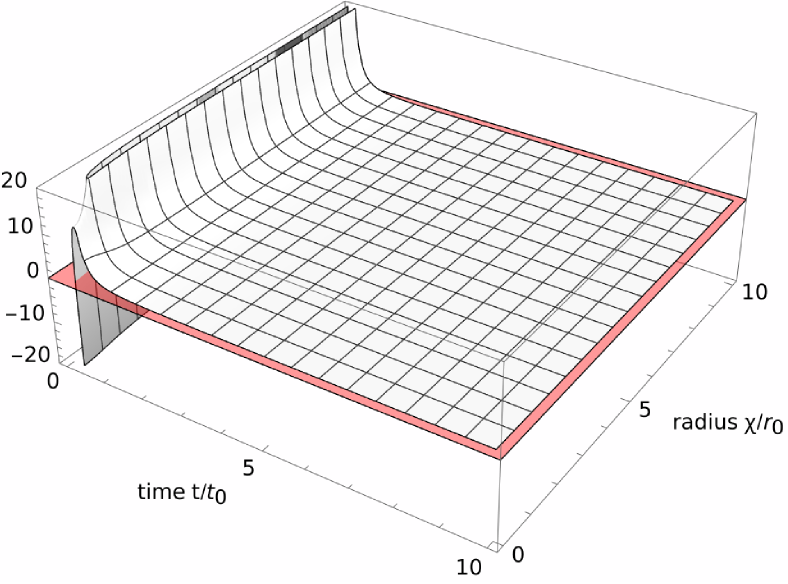}
    \caption{Plot of $\rho - \left| p^\mathrm{angular} \right|$ normalised by $\frac{{H_0}^2}{\kappa^2}$ with $r_0 H_0 = 10^{0}$, $w=-2/3$}
    \label{fig4:sub6}
\end{subfigure}
\caption{The behaviours of $\rho$ \eqref{fig4:sub1}, $\rho + p^\mathrm{radial} $ \eqref{fig4:sub2}, $\rho + p^\mathrm{angular}$ \eqref{fig4:sub3}, 
$\rho + p^\mathrm{radial} + 2 p^\mathrm{angular} $ \eqref{fig4:sub4}, 
$\rho - \left| p^\mathrm{radial}\right|$ \eqref{fig4:sub5} and $\rho - \left| p^\mathrm{angular} \right|$ \eqref{fig4:sub6}, 
when we use the normalisation of $\frac{{H_0}^2}{\kappa^2}$ 
with $r_0 H_0 = 10^{0}$, $w=-2/3$. 
The red plane indicates $z=0$, and each quantity takes positive values above this plane.
All the energy conditions are violated only in the early time, although the SEC is always violated.}
\label{fig4:overall}
\end{figure}

Figs.~\ref{fig2:overall} and \ref{fig4:overall} are those six quantities for $w=-2/3$.
In Fig~\ref{fig2:overall} with $r_0 H_0 = 10^{-2}$ input, all the energy conditions except SEC are violated, but they all might be recovered in the late time, as in Fig.~\ref{fig1:overall}.
The SEC is always violated.
On the other hand, in Fig.~\ref{fig4:overall} with $r_0 H_0 = 10^{0}$ input, all the energy conditions are violated only in the early time, although the SEC is still violated. 
Therefore, when the throat radius is larger than or equal to the Hubble radius, the energy conditions could be recovered except SEC.

\subsection{Case 2: $w=-1$}

\begin{figure}[htbp]
\centering
\begin{subfigure}[b]{0.3\textwidth}
    \centering
    \includegraphics[width=\textwidth]{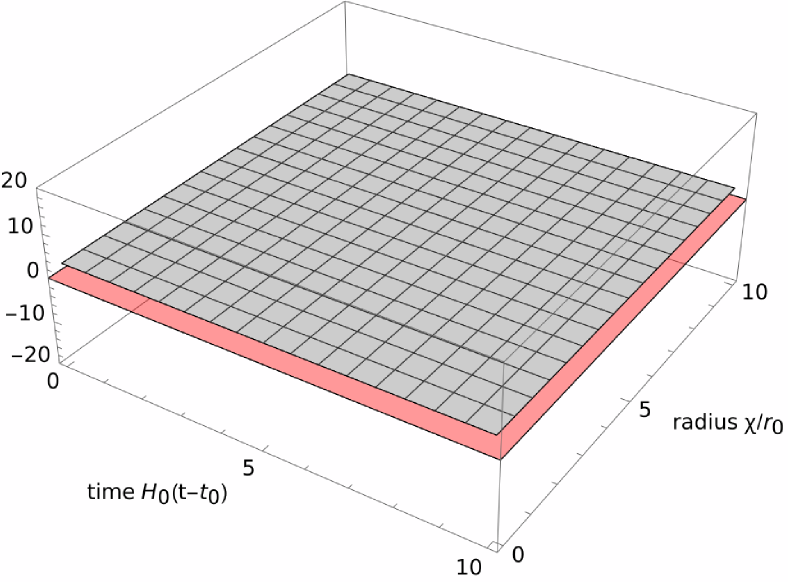}
    \caption{Plot of $\rho$ normalised by $\frac{{H_0}^2}{\kappa^2}$ with $r_0 H_0 = 10^{-2}$, $w=-1$}
    \label{fig5:sub1}
\end{subfigure}
\hfill
\begin{subfigure}[b]{0.3\textwidth}
    \centering
    \includegraphics[width=\textwidth]{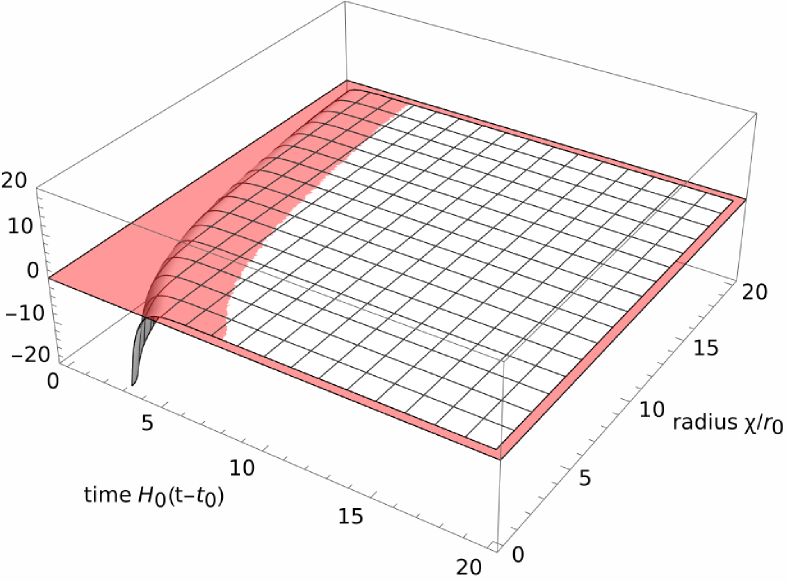}
    \caption{Plot of $\rho + p^\mathrm{radial} $ normalised by $\frac{{H_0}^2}{\kappa^2}$ with $r_0 H_0 = 10^{-2}$, $w=-1$}
    \label{fig5:sub2}
\end{subfigure}
\hfill
\begin{subfigure}[b]{0.3\textwidth}
    \centering
    \includegraphics[width=\textwidth]{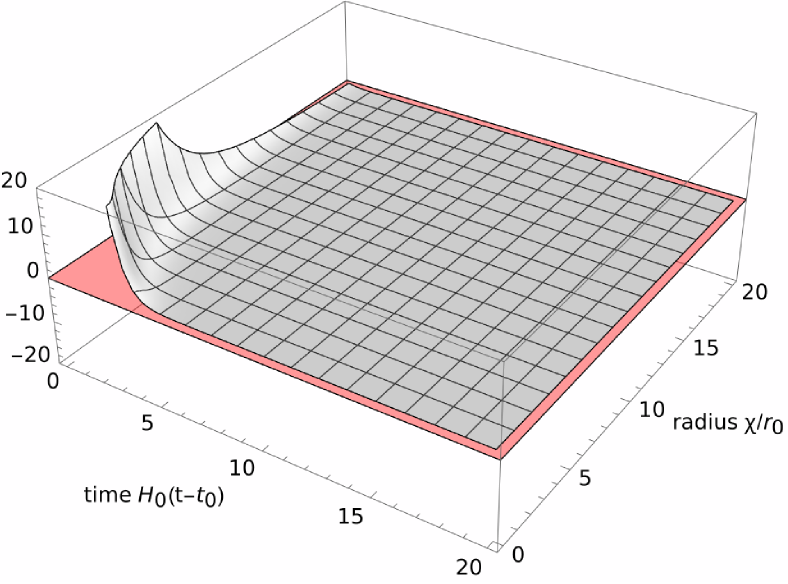}
    \caption{Plot of $\rho+p^\mathrm{angular} $ normalised by $\frac{{H_0}^2}{\kappa^2}$ with $r_0 H_0 = 10^{-2}$, $w=-1$}
    \label{fig5:sub3}
\end{subfigure}
\hfill
\begin{subfigure}[b]{0.3\textwidth}
    \centering
    \includegraphics[width=\textwidth]{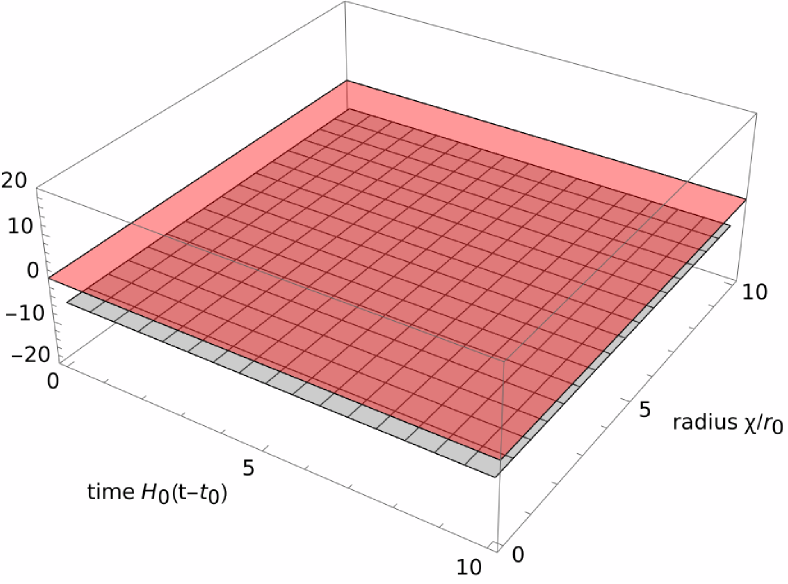}
    \caption{Plot of $\rho + p^\mathrm{radial} + 2 p^\mathrm{angular} $ normalised by $\frac{{H_0}^2}{\kappa^2}$ with $r_0 H_0 = 10^{-2}$, $w=-1$}
    \label{fig5:sub4}
\end{subfigure}
\hfill
\begin{subfigure}[b]{0.3\textwidth}
    \centering
    \includegraphics[width=\textwidth]{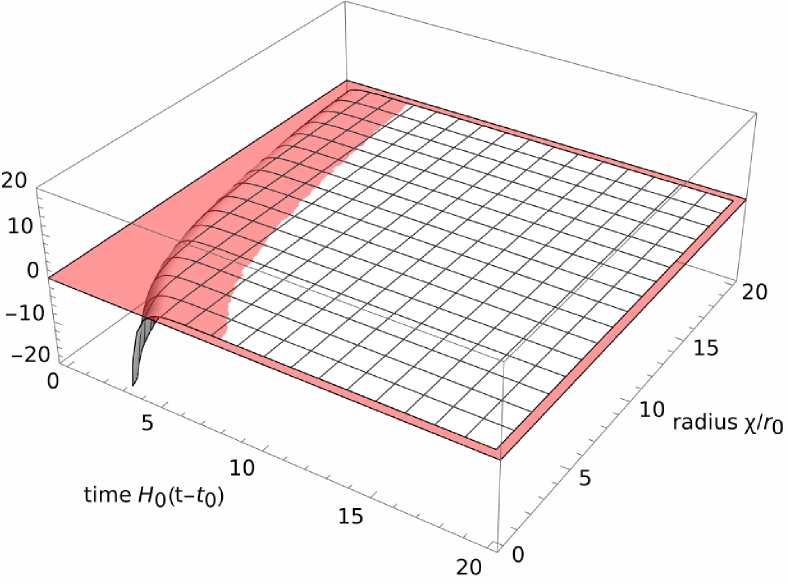}
    \caption{Plot of $\rho - \left| p^\mathrm{radial} \right|$ normalised by $\frac{{H_0}^2}{\kappa^2}$ with $r_0 H_0 = 10^{-2}$, $w=-1$}
    \label{fig5:sub5}
\end{subfigure}
\hfill
\begin{subfigure}[b]{0.3\textwidth}
    \centering
    \includegraphics[width=\textwidth]{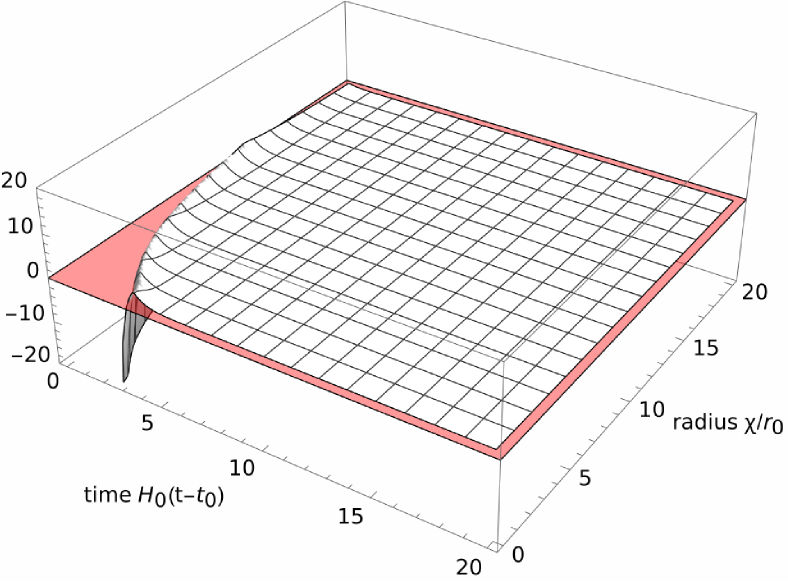}
    \caption{Plot of $\rho - \left| p^\mathrm{angular} \right|$ normalised by $\frac{{H_0}^2}{\kappa^2}$ with $r_0 H_0 = 10^{-2}$, $w=-1$}
    \label{fig5:sub6}
\end{subfigure}
\caption{The behaviours of $\rho$ \eqref{fig5:sub1}, $\rho + p^\mathrm{radial} $ \eqref{fig5:sub2}, $\rho + p^\mathrm{angular}$ \eqref{fig5:sub3}, 
$\rho + p^\mathrm{radial} + 2 p^\mathrm{angular} $ \eqref{fig5:sub4}, 
$\rho - \left| p^\mathrm{radial}\right|$ \eqref{fig5:sub5} and $\rho - \left| p^\mathrm{angular} \right|$ \eqref{fig5:sub6}, 
when we use the normalisation of $\frac{{H_0}^2}{\kappa^2}$ 
with $r_0 H_0 = 10^{-2}$, $w=-1$. 
The red plane indicates $z=0$, and each quantity takes positive values above this plane.
Although all the energy conditions are violated, they are all recovered in the late time, except SEC, which is always violated.}
\label{fig5:overall}
\end{figure}

In the case $w= -1$, Eq.~\eqref{EC2} yields
\begin{align}
\label{EC case2}
    \rho = \frac{3{H_0}^2}{\kappa^2} 
    \, , \quad 
    \rho + p^\mathrm{radial} 
    &=
    - \frac{\e^{-2H_0 (t-t_0)}}{\kappa^2 {r_0}^2\left[1+ \left(\frac{\chi}{r_0}\right)^2\right]^\frac{3}{2}} 
    \, , \quad 
    \rho + p^\mathrm{angular} 
    = 
    \frac{\e^{-2H_0 (t-t_0)}}{2\kappa^2 {r_0}^2\left[1+ \left(\frac{\chi}{r_0}\right)^2\right]^\frac{3}{2}} 
    \, , \nonumber \\
    \rho + p^\mathrm{radial} + 2 p^\mathrm{angular} 
    &=
    - \frac{6 {H_0}^2}{\kappa^2} 
    \, , \nonumber \\
    \rho - \left| p^\mathrm{radial} \right| 
    &=
    \frac{{H_0}^2}{\kappa^2} \left[ 3 - \left| 3 + \frac{1}{(r_0 H_0)^2} \frac{1}{\left[1+ \left(\frac{\chi}{r_0}\right)^2\right]^\frac{3}{2}} 
    \e^{-2H_0 (t-t_0)} \right| \right] 
    \nonumber \\
    \rho - \left| p^\mathrm{angular} \right| 
    &=
    \frac{{H_0}^2}{\kappa^2} \left[ 3 - \left| 3 - \frac{1}{2} \frac{1}{(r_0 H_0)^2} \frac{1}{\left[1+ \left(\frac{\chi}{r_0}\right)^2\right]^\frac{3}{2}} 
    \e^{-2H_0 (t-t_0)} \right| \right] 
    \, .
\end{align}
We plot the above six quantities as functions of $(H_0 (t-t_0), \chi/r_0)$.
Fig.~\ref{fig5:overall} shows $\rho$, $\rho + p^\mathrm{radial} $, $\rho + p^\mathrm{angular}$, $\rho + p^\mathrm{radial} + 2 p^\mathrm{angular} $, $\rho - \left| p^\mathrm{radial}\right|$ and $\rho - \left| p^\mathrm{angular} \right|$ 
for $r_0 H_0 = 10^{-2}$. 
All the energy conditions are violated, but they are all recovered in the late time, except SEC, which is always violated.

\subsection{Case 3: $w<-1$}

\begin{figure}[htbp]
\centering
\begin{subfigure}[b]{0.3\textwidth}
    \centering
    \includegraphics[width=\textwidth]{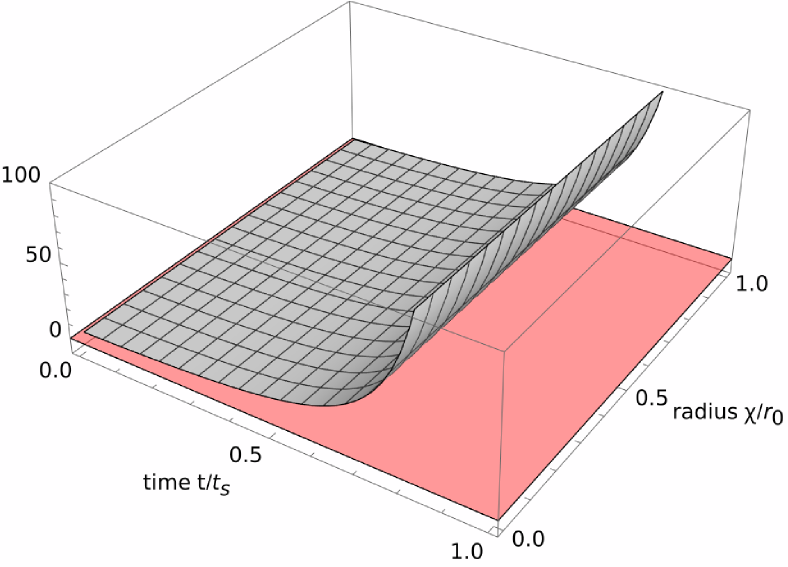}
    \caption{Plot of $\rho$ normalised by $\frac{{H_0}^2}{\kappa^2}$ with $r_0 H_0 = 10^{-2}$, $w=-4/3$}
    \label{fig6:sub1}
\end{subfigure}
\hfill
\begin{subfigure}[b]{0.3\textwidth}
    \centering
    \includegraphics[width=\textwidth]{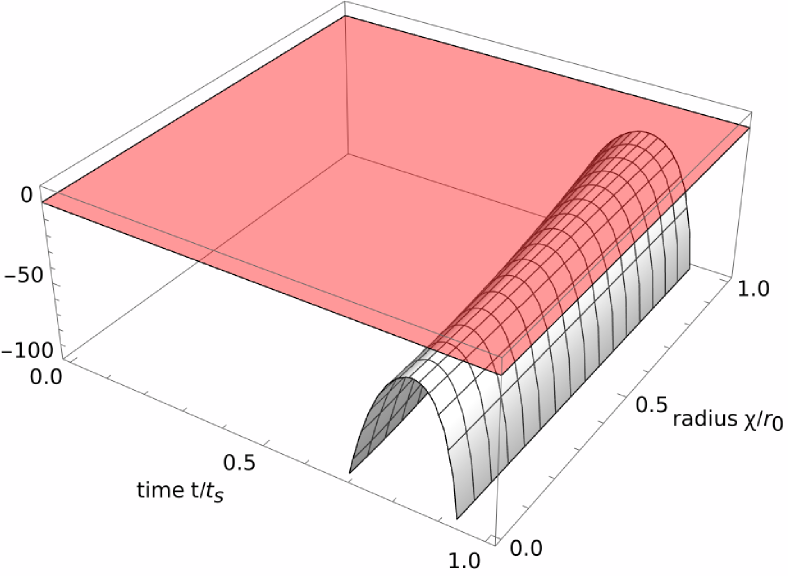}
    \caption{Plot of $\rho + p^\mathrm{radial} $ normalised by $\frac{{H_0}^2}{\kappa^2}$ with $r_0 H_0 = 10^{-2}$, $w=-4/3$}
    \label{fig6:sub2}
\end{subfigure}
\hfill
\begin{subfigure}[b]{0.3\textwidth}
    \centering
    \includegraphics[width=\textwidth]{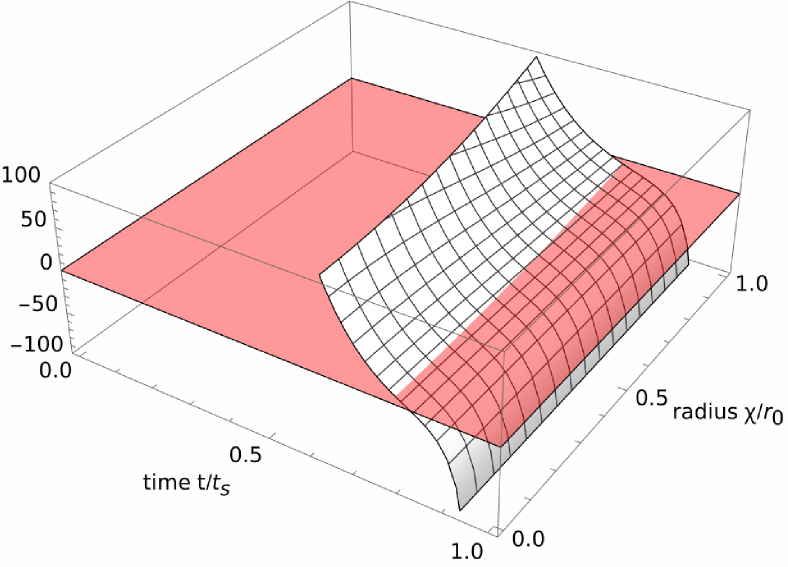}
    \caption{Plot of $\rho+p^\mathrm{angular} $ normalised by $\frac{{H_0}^2}{\kappa^2}$ with $r_0 H_0 = 10^{-2}$, $w=-4/3$}
    \label{fig6:sub3}
\end{subfigure}
\hfill
\begin{subfigure}[b]{0.3\textwidth}
    \centering
    \includegraphics[width=\textwidth]{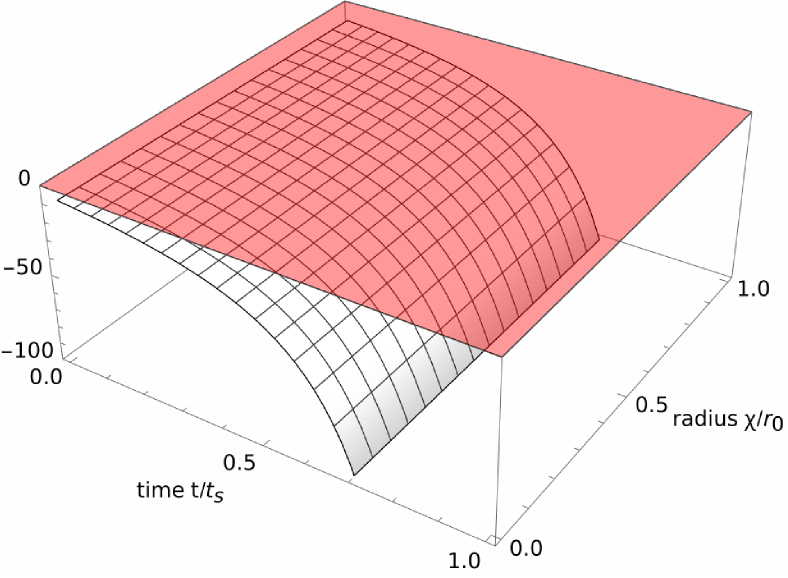}
    \caption{Plot of $\rho + p^\mathrm{radial} + 2 p^\mathrm{angular} $ normalised by $\frac{{H_0}^2}{\kappa^2}$ with $r_0 H_0 = 10^{-2}$, $w=-4/3$}
    \label{fig6:sub4}
\end{subfigure}
\hfill
\begin{subfigure}[b]{0.3\textwidth}
    \centering
    \includegraphics[width=\textwidth]{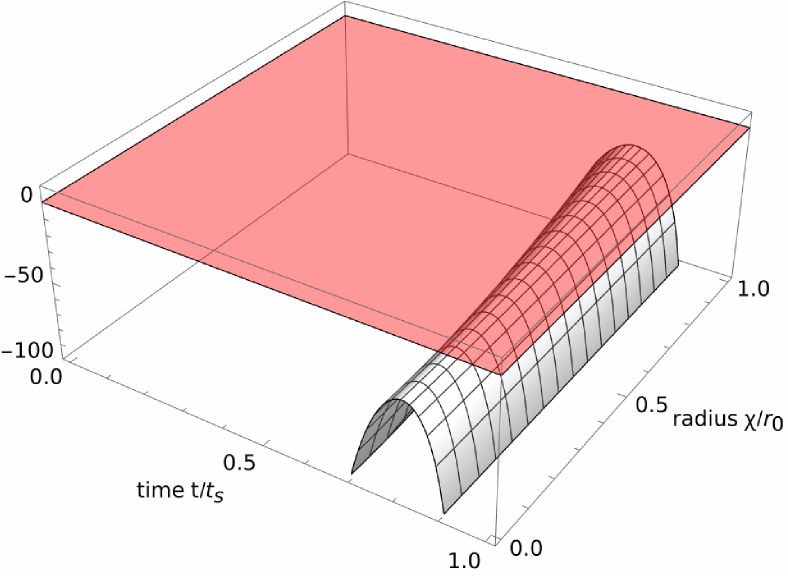}
    \caption{Plot of $\rho - \left| p^\mathrm{radial} \right|$ normalised by $\frac{{H_0}^2}{\kappa^2}$ with $r_0 H_0 = 10^{-2}$, $w=-4/3$}
    \label{fig6:sub5}
\end{subfigure}
\hfill
\begin{subfigure}[b]{0.3\textwidth}
    \centering
    \includegraphics[width=\textwidth]{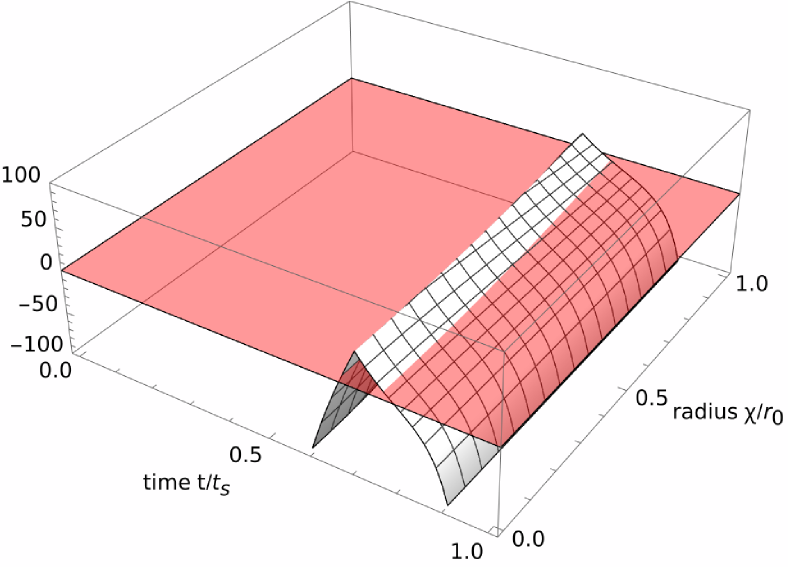}
    \caption{Plot of $\rho - \left| p^\mathrm{angular} \right|$ normalised by $\frac{{H_0}^2}{\kappa^2}$ with $r_0 H_0 = 10^{-2}$, $w=-4/3$}
    \label{fig6:sub6}
\end{subfigure}
\caption{The behaviours of $\rho$ \eqref{fig6:sub1}, $\rho + p^\mathrm{radial} $ \eqref{fig6:sub2}, $\rho + p^\mathrm{angular}$ \eqref{fig6:sub3}, 
$\rho + p^\mathrm{radial} + 2 p^\mathrm{angular} $ \eqref{fig6:sub4}, 
$\rho - \left| p^\mathrm{radial}\right|$ \eqref{fig1:sub5} and $\rho - \left| p^\mathrm{angular} \right|$ \eqref{fig1:sub6}, 
when we use the normalisation of $\frac{{H_0}^2}{\kappa^2}$ 
with $r_0 H_0 = 10^{-2}$, $w=-4/3$. 
All the energy conditions are always violated, as in the standard phantom universe.}
\label{fig6:overall}
\end{figure}

In the case $w<-1$, Eq.~\eqref{EC2} yields
\begin{align}
\label{EC case3}
    \rho 
    &= \frac{3{H_0}^2}{\kappa^2} \left(1 - \frac{t}{t_s} \right)^{-2} 
    \, , \nonumber \\
    \rho + p^\mathrm{radial} 
    &=
    \frac{{H_0}^2 }{\kappa^2} \left[ \frac{3(1+w)}{2}  \left( 1 - \frac{t}{t_s} \right)^{-2}
    - \frac{1}{(r_0 H_0)^2} 
    \frac{1}{\left[1+ \left(\frac{\chi}{r_0}\right)^2\right]^\frac{3}{2}} \left( 1 - \frac{t}{t_s} \right)^{-\frac{4}{3(1+w)}} 
    \right] 
    \, , \nonumber \\
    \rho + p^\mathrm{angular} 
    &=
    \frac{{H_0}^2}{\kappa^2} \left[ \frac{3(1+w)}{2}   \left( 1 - \frac{t}{t_s} \right)^{-2}
    + \frac{1}{2} \frac{1}{(r_0 H_0)^2} 
    \frac{1}{\left[1+ \left(\frac{\chi}{r_0}\right)^2\right]^\frac{3}{2}} \left( 1 - \frac{t}{t_s}\right)^{-\frac{4}{3(1+w)}} 
    \right] 
    \, , \nonumber \\
    \rho + p^\mathrm{radial} + 2 p^\mathrm{angular} 
    &=
    \frac{3{H_0}^2(1+3w)}{\kappa^2} 
    \left( 1 - \frac{t}{t_s} \right)^{-2} 
    \, , \nonumber \\
    \rho - \left| p^\mathrm{radial} \right| 
    &=
    \frac{{H_0}^2}{\kappa^2} \left[ 3 \left( 1 - \frac{t}{t_s} \right)^{-2} - \left| 3w  \left( 1 - \frac{t}{t_s}  \right)^{-2} 
    - \frac{1}{(r_0 H_0)^2} \frac{1}{\left[1+ \left(\frac{\chi}{r_0}\right)^2\right]^\frac{3}{2}} \left( 1 - \frac{t}{t_s}  \right)^{-\frac{4}{3(1+w)}} \right| \right] 
    \, , \nonumber \\
    \rho - \left| p^\mathrm{angular} \right| 
    &=
    \frac{{H_0}^2}{\kappa^2} \left[ 3 \left( 1 - \frac{t}{t_s}  \right)^{-2} - \left| 3w \left( 1 - \frac{t}{t_s} \right)^{-2} 
    + \frac{1}{2} \frac{1}{(r_0 H_0)^2} \frac{1}{\left[1+ \left(\frac{\chi}{r_0}\right)^2\right]^\frac{3}{2}} \left( 1 - \frac{t}{t_s}  \right)^{-\frac{4}{3(1+w)}} \right| \right] 
    \, .
\end{align}
Here, we input $t_0 = 0$. 
We plot the above six quantities as functions of $(t/t_s, \chi/r_0)$.
Fig.~\ref{fig6:overall} shows $\rho$, $\rho + p^\mathrm{radial} $, $\rho + p^\mathrm{angular}$, $\rho + p^\mathrm{radial} + 2 p^\mathrm{angular}$, $\rho - \left| p^\mathrm{radial}\right|$ and $\rho - \left| p^\mathrm{angular} \right|$ for $r_0 H_0 = 10^{-2}$ and $w=-4/3$. 
All the energy conditions are always violated, as in the standard phantom universe.

\subsection{Recovery of energy conditions}

In the previous three subsections, we considered embedding the wormhole into the accelerated expansion of the Universe for three representative cases: quintessence, a cosmological constant, and phantom energy. 
We find that the contribution of these DE components enters the total energy–momentum tensor through an effective cosmological fluid, whose dominance depends on the relative scales of the wormhole $r_0$ and the background expansion $H_0$.

For quintessence and the cosmological constant, we find that the energy conditions can be recovered when the DE component becomes dominant in the total energy budget. 
This corresponds to a regime in which the repulsive effect associated with cosmic acceleration sufficiently contributes to the overall energy and pressure, compensating for the violations that typically arise from the wormhole sector. 
An important outcome of this analysis is that the recovery of the energy conditions is not universal, but scale-dependent. 
The energy conditions are evaluated in terms of the ratio of the wormhole radius to the Hubble radius $r_0 H_0$.
When the wormhole size is much smaller than the Hubble radius $r_0 H_0 \ll1$, the local gravitational effects associated with the wormhole geometry become relatively stronger, making it more difficult for the cosmic expansion to restore the energy conditions.

In contrast, for the phantom case, the situation is qualitatively different. 
Since phantom energy intrinsically violates several energy conditions, the inclusion of this component does not lead to an overall recovery of the energy conditions, even when cosmic expansion is taken into account. 
This distinction highlights that the recovery mechanism is not a generic consequence of cosmic expansion, but rather depends on the physical nature of the cosmological fluid.

\section{Singular spacetime with wormhole}
\label{SecVIII}

In the previous section, we considered the wormhole spacetime with $N=N(\tau)$.
In this section, at last, we discuss the wormhole spacetime with a singularity with $N=N(\tau, \chi)$ including the spatial dependence. 
However, we assume that in the limit of $\chi\to +\infty$, $N(\tau,\chi)$ is a finite function of $\tau$ except for discrete singular points which correspond to future or past singularity.  
We define a time coordinate $t$ asymptotically by 
\begin{align}
    dt = \e^{\lim_{\chi\to +\infty} N(\tau, \chi)} d\tau
    \, ,
\end{align}
and $t$ can be identified with the cosmic time in the FLRW universe.
With this asymptotic identification, the relation with future singularities becomes manifest.

In the following, we demonstrate the embedding of the wormhole into the spacetime, showing a future singularity, in two particular examples.
Note that the finite-time future singularity arises from the effective DE behavior induced by the scalar-field sector, rather than from the wormhole geometry itself.
That is, the wormhole solution is embedded in a cosmological background driven independently by the dark-energy component arising from the four-scalar fields. 
We illustrate cosmological scenarios where both phenomena can naturally emerge.

\subsection{Example 1}

As a first example, we consider the following. 
\begin{align}
\label{ex1}
    N\left( \tau (t), \chi \right) 
    = 
    N_0 \left[
    \frac{\left( t_0 - t \right)^2}{{t_1}^2} - \frac{\chi_0 \chi}{{\chi_0}^2 + \chi^2} 
    \right]^\frac{\beta +1}{2}
    \, ,
\end{align}
Here, $N_0$, $t_0$, $t_1$, $\chi_0$ and $\beta$ are constants, and we assume $N_0$, $t_1$ and $\chi_0$ are positive. 
If $\beta$ is not an integer greater or equal to $-1$, there appear the singularities when 
\begin{align}
\label{ex1sing}
    t = t_0 \pm t_1 \sqrt{\frac{\chi_0 \chi}{{\chi_0}^2 + \chi^2}}\, .
\end{align}
When $\chi$ is negative, there is no real solution, and therefore, there is no singularity in another universe. 
On the other hand, there appears a singularity in our universe where $l$ is positive. 
When $\chi\to \infty$,  we find 
\begin{align}
\label{Nbeta}
    N(\tau,\chi) 
    \equiv 
    N\left( \tau(t), \chi \right) 
    \to 
    \frac{N_0}{{t_1}^{\beta+1}} \left( t_0 - t \right)^{\beta +1}\, . 
\end{align} 
For the Hubble parameter $H=\frac{d}{dt} N\left( \tau (t), \chi\to +\infty\right)$, comparing Eq.~\eqref{Nbeta} with Eq.~\eqref{Halpha} or \eqref{HalphaB}, we find $\alpha=\beta$. 
Therefore, Type I and Type III singularity appear for $\beta\leq -1$ and $-1<\beta<0$ respectively.
Furthermore, Type II singularity appears for $0<\beta<1$,
and Type IV singularity appears if $\beta>1$ but $\beta$ is not an integer. 

\begin{figure}
\begin{center}
\unitlength=1mm
\begin{picture}(120,120)

\put(10,60){\line(1,1){50}}
\put(10,60){\line(1,-1){50}}
\put(110,60){\line(-1,1){50}}
\put(110,60){\line(-1,-1){50}}

\put(100,91){\makebox(0,0){future null infinity}}
\put(100,89){\makebox(0,0){in our universe}}

\put(100,31){\makebox(0,0){past null infinity}}
\put(100,29){\makebox(0,0){in our universe}}

\put(20,91){\makebox(0,0){future null infinity}}
\put(20,89){\makebox(0,0){in another universe}}

\put(20,31){\makebox(0,0){past null infinity}}
\put(20,29){\makebox(0,0){in another universe}}

\put(50,41){\makebox(0,0){throat}}
\put(50,39){\makebox(0,0){of wormhole}}

\put(75,80){\makebox(0,0){future singularity}}

\thicklines

\put(60,10){\line(0,1){100}}

\qbezier(60,80)(60,81)(61,82)
\qbezier(61,82)(62,83)(63,82)
\qbezier(63,82)(64,81)(65,82)
\qbezier(65,82)(66,83)(67,82)
\qbezier(67,82)(68,81)(69,82)
\qbezier(69,82)(70,83)(71,82)
\qbezier(71,82)(72,81)(73,82)
\qbezier(73,82)(74,83)(75,82)
\qbezier(75,82)(76,81)(77,82)
\qbezier(77,82)(78,83)(79,82)
\qbezier(79,82)(80,81)(81,82)
\qbezier(81,82)(82,83)(83,82)
\qbezier(83,82)(84,81)(85,82)
\qbezier(85,82)(86,83)(90,80)

\qbezier(60,80)(60,79)(61,78)
\qbezier(61,78)(62,77)(63,78)
\qbezier(63,78)(64,79)(65,78)
\qbezier(65,78)(66,77)(67,78)
\qbezier(67,78)(68,79)(69,78)
\qbezier(69,78)(70,77)(71,78)
\qbezier(71,78)(72,79)(73,78)
\qbezier(73,78)(74,77)(75,78)
\qbezier(75,78)(76,79)(77,78)
\qbezier(77,78)(78,77)(79,78)
\qbezier(79,78)(80,79)(81,78)
\qbezier(81,78)(82,77)(83,78)
\qbezier(83,78)(84,79)(85,78)
\qbezier(85,78)(86,77)(90,80)

\qbezier(70,60)(55,75)(55,80)
\qbezier(55,80)(55,85)(65,95)

\put(55,80){\makebox(0,0){\vector(0,1){0}}}
\put(65,95){\vector(1,1){1}}

\end{picture}
\end{center}
\caption{The Penrose diagram of a wormhole \eqref{ex1} with a future singularity in our universe. 
Because the singularity occurs in the space-like hypersurface except the region near the throat of the wormhole $l\sim 0$, 
we can consider the realistic time-like orbit, where a particle goes to another universe before the singularity 
through the wormhole and returns to our universe after the singularity. 
As shown by a curve with arrows, we can avoid the singularity in our universe by escaping 
from our universe to another universe by going through the wormhole, and we can return to our universe after the singularity. 
}
\label{FigWHS}
\end{figure}
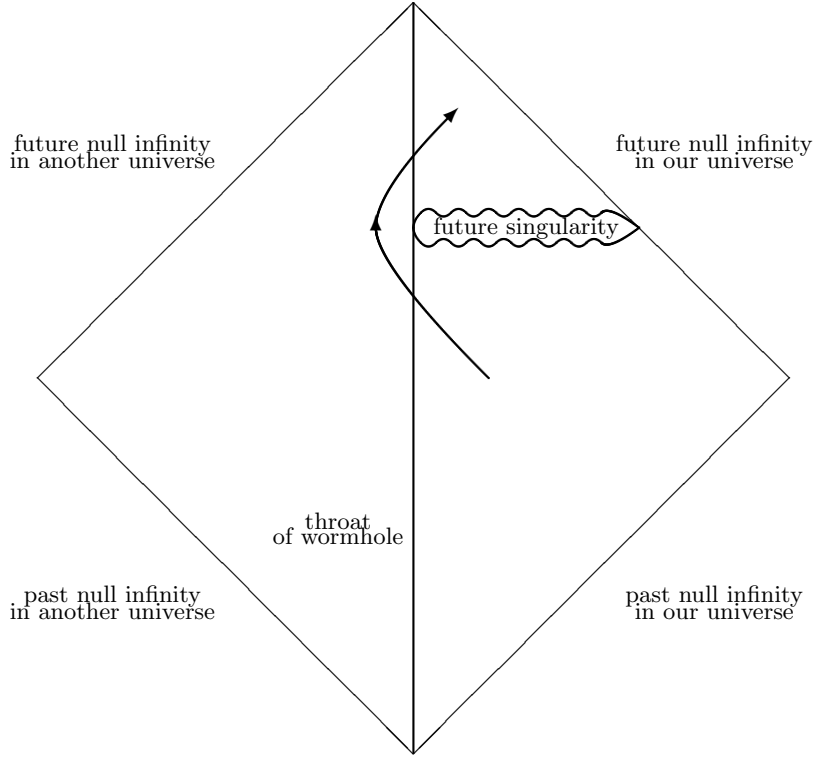

The singularity shows up in the space-like hypersurface except in the region near the throat of the wormhole $l\sim 0$.
We can consider a realistic time-like orbit in which a particle travels to another universe through the wormhole before the singularity.
As depicted in Fig.~\ref{FigWHS}, in our model, because the future singularity does not appear in another universe, we can avoid the singularity in our universe by escaping 
from our universe to another universe by going through the wormhole, and we can return to our universe after the singularity.

\subsection{Example 2}

We also consider the following second example, 
\begin{align}
\label{ex2}
    N\left( \tau(t), \chi \right) 
    = 
    N_0 \left[ 
    \frac{\left( t_0 - t \right)^2}{{t_1}^2} +  1 - \tanh \left( \frac{\chi}{\chi_1} \right) 
    \right]^\frac{\beta +1}{2}
    \, ,
\end{align}
Here $N_0$, $t_0$, $t_1$, $\chi_1$ and $\beta$ are constants. 
We assume $N_0$, $t_1$ and $\chi_1$ are positive. 
When $\chi\to + \infty$, $N$ behaves as in Eq.~\eqref{Nbeta}, and the future singularity could appear.
However, for finite $\chi$ and in the limit of $\chi\to -\infty$, there is no singularity, although there could occur the rips of any extended objects for positive and large $\chi$, $\chi\gg \chi_1$. 
In this sense, there is no real future singularity in the model Eq.~\eqref{ex2}.

\section{Wormhole motivated by Dark Matter Halo}
\label{SecIX}

In previous sections, we have studied the wormhole geometry embedded into the FLRW universe and analysed the energy conditions in such a spacetime.
However, it is possible to embed a wormhole geometry into another spacetime.
There was recently some interest in wormholes corrected by the density profiles of dark matter (DM) halos in galaxies.
For instance, Ref.~\cite{Errehymy:2025vvs} is devoted to wormhole geometry corrected by the Navarro-Frenk-White (NFW) profile~\cite{Navarro:1995iw, Navarro:1996gj} and other profiles~\cite{Begeman:1991iy, Boehmer:2007um, Capolupo:2024ckb}. 
In Ref.~\cite{Errehymy:2025vvs}, the wormhole is assumed to alternate the supermassive black hole at the galactic centre. 
Several other kinds of wormhole spacetime with DM profile accounts have been proposed (see~\cite{Rahaman:2023tkm}).
In this section, we briefly review such a wormhole geometry corrected by a DM profile. 
We demonstrate that such a wormhole may be naturally realised within our four-scalar framework. 
In other words, the four-scalar framework gives the same effect as DM profile to a wormhole without any account of real DM.
In this section, we briefly review a wormhole geometry in \cite{Rahaman:2023tkm} as one example and examine it within our four-scalar framework.

In the NFW profile, the energy density $\rho(r)$ has the following form, 
\begin{align}
\label{NFWprofile}
    \rho = \rho_\mathrm{NFW} (r) \equiv \frac{\rho_s {R_s}^3}{r\left( r + R_s \right)^2}\, .
\end{align}
$R_s$ is the characteristic scale of the cold DM halo radius, and $\rho_s$ is used for the normalisation of the density. 
For instance, $\rho_s = 0.008\times 10^{7.5}\, M_\odot \mathrm{kpc}^{-3}$ and $R_s = 130\,\mathrm{kpc}$ for the M87 galaxy. 
The NFW profile is empirically obtained by the analysis of an $N$-body simulation based on Newtonian gravity, as other DM halo profiles are also obtained empirically or theoretically.
The NFW profile has a singularity at the centre $r=0$, 
and it cannot be applied for small $r$ or in the region where we cannot ignore the curvature in the spacetime. 
Here, one important possibility is that there might be a wormhole instead of a black hole at the galactic centre.

Although most of the DM halo profiles do not follow from general relativity, the following wormhole geometry has been proposed~\cite{Errehymy:2025vvs}: 
\begin{align}
\label{DMWH1}
    ds^2 
    &= 
    \e^{2\Phi(r)} dt^2 + \frac{dr^2}{1 - \frac{\epsilon(r)}{r}}
    + r^2 \left[ d\theta^2 + \sin^2 \theta \left( d \phi - \omega(r) dt \right)^2 \right]
    \, . 
\end{align}
$\Phi(r)$ is called the redshift function, and $\omega(r)$ represents the angular velocity introduced to describe a slowly rotating wormhole.
$\epsilon(r)$ is called the shape function and determined by solving the $(t,t)$ component of the Einstein equation with $\omega=0$ for the given DM halo profile, 
\begin{align}
\label{epsilon}
    \epsilon'(r) = 8\pi r^2 \rho (r) 
    \, .
\end{align}
For the NFW profile, we find 
\begin{align}
\label{NFWepsilon2}
\epsilon(r) = \epsilon_\mathrm{NFW}(r) \equiv 8\pi \rho_s {R_s}^3 \left[ \ln \left( r + R_s \right) 
+ \frac{R_s}{r+R_s} \right] + \epsilon_0\, .
\end{align}
Here, the integration constant $\epsilon_0$ is determined so that $\epsilon_\mathrm{NFW}\left( r_0 \right) = r_0$, where $r_0$ is the minimal radius of the wormhole throat. 

The redshift function $\Phi$ can be determined if the $(r,r)$ component of the Einstein equation is consistent with the $(\theta,\theta)$ and $(\phi,\phi)$ components. 
Or, for the given redshift function $\Phi$, the pressures in the radial and angular directions can be determined by the Einstein equation, as we will see later. 
We note that non-vanishing pressures generally appear; therefore, the DM considered in Ref.~\cite{Errehymy:2025vvs} cannot be regarded as ordinary cold (pressureless) dark matter, but rather as a similar counterfeit.
In other words, a fluid corresponding to such a DM profile cannot be realised in ordinary ways, but the wormhole geometry as in Eq.~\eqref{DMWH1} can be stably realised by the four-scalar model in Eq.~\eqref{Eq. acg1}.

For the geometry~\eqref{DMWH1}, the non-vanishing Christoffel symbols have the following expressions, 
\begin{align}
\label{Chrstfl}
    \Gamma^{t}_{r t} 
    &= 
    \Phi^{\prime} 
    -\frac{1}{2} r^2 \sin^2\theta \, \e^{-2 \Phi} \omega \omega^{\prime} 
    \, , \quad 
    \Gamma^{t}_{\phi r} 
    = 
    \frac{1}{2} r^2 \sin^2\theta \, \e^{-2 \Phi} \omega^{\prime} 
    \, , \nonumber \\
    \Gamma^{r}_{t t} 
    &=
    \frac{1}{r} 
    (r-\epsilon) 
    \left[ \e^{2 \Phi} \Phi^{\prime}-r \sin^2\theta  \omega \left(r \omega^{\prime}+\omega\right)\right]
    \, , \quad 
    \Gamma^{r}_{r r} 
    = 
    - \frac{1}{2r}\frac{\epsilon-r \epsilon^{\prime}}{r-\epsilon} 
    \, , \nonumber \\
    \Gamma^{r}_{\theta \theta} 
    &=
    \epsilon-r \, , \quad 
    \Gamma^{r}_{\phi t} 
    = 
    \frac{1}{2} \sin^2\theta  (r-\epsilon) \left(r \omega^{\prime}+2 \omega\right) 
    \, , \quad 
    \Gamma^{r}_{\phi \phi} 
    = 
    \sin^2\theta  \left( \epsilon-r \right) 
    \, , \nonumber \\
    \Gamma^{\theta}_{t t} 
    &=
    \, - \sin\theta \cos\theta \omega^2 
    \, , \quad 
    \Gamma^{\theta}_{\theta r} 
    = 
    \frac{1}{r} \, , \quad 
    \Gamma^{\theta}_{\phi t} 
    =  
    \sin\theta  \cos\theta  \omega 
    \, , \quad 
    \Gamma^{\theta}_{\phi \phi} 
    =
    -\sin\theta  \cos\theta 
    \, , \nonumber \\
    \Gamma^{\phi}_{r t} 
    &= 
    - \frac{1}{2 r}
    \left[ 
    r^3 \sin^2\theta  \e^{-2 \Phi} \omega^2 \omega^{\prime} 
    + 2\omega \left(1-r \Phi^{\prime}\right)+r \omega^{\prime}
    \right]
    \, , \quad 
    \Gamma^{\phi}_{\theta t} 
    =
    -\cot \theta \omega 
    \, , \nonumber \\
    \Gamma^{\phi}_{\phi r} 
    &=
    \frac{1}{2} r^2 \sin^2\theta  \e^{-2 \Phi} \omega \omega^{\prime}+\frac{1}{r} 
    \, , \quad 
    \Gamma^{\phi}_{\phi \theta} 
    = 
    \cot \theta 
    \, .
\end{align}
Then, the non-vanishing components of the Ricci tensor and the Ricci scalar are given by 
\begin{align}
\label{Ricci}
    R_{t t} 
    &=
    \frac{1}{2} r^3 \sin^4 \theta \, \e^{-2 \Phi} \omega^2 (\epsilon-r) (\omega^{\prime})^{2} 
    \nonumber \\
    & 
    \quad \ 
    + \frac{1}{2 r} \sin^2\theta 
    \left\{ 
    r \left[ 
    -r^2 (\omega^{\prime})^{2} 
    +r \omega 
    \left(\omega ^{\prime} \left(2 r \Phi^{\prime}+\epsilon^{\prime}-8\right)-2 r \omega^{\prime \prime}\right) 
    +\omega^2 \left(\epsilon^{\prime}-2 r \Phi^{\prime}\right) 
    \right] 
    \right.
    \nonumber \\
    & 
    \left. \qquad \qquad \qquad \qquad 
    +\epsilon 
    \left[
    r^2 (\omega^{\prime})^{2}+r \omega \left(\left(7-2 r \Phi^{\prime}\right) \omega ^{\prime}+2 r \omega^{\prime \prime}\right)
    +\omega^2 \left(2 r \Phi^{\prime}+1\right)
    \right] 
    \right\} 
    \nonumber \\
    &
    \quad \ 
    + \frac{\e^{2 \Phi} }{2 r^2} 
    \left\{ 
    \Phi^{\prime} 
    \left[- r \epsilon^{\prime} +2 r (r-\epsilon) \Phi^{\prime}-3 \epsilon+4 r\right] 
    +2 r (r-\epsilon) \Phi^{\prime \prime} 
    \right\} 
    \, , \nonumber \\
    R_{r r} 
    &=
    \frac{r^2}{2} \sin^2\theta \ \e^{-2 \Phi} (\omega^{\prime})^{2} 
    \nonumber \\
    & \quad \ 
    +\frac{1}{2 r^2 (r-\epsilon)} 
    \left\{ 
    -2 r^3 \left[\Phi^{\prime \prime}+(\Phi^{\prime})^2\right] +r \left( r \Phi^{\prime}+2\right) \epsilon^{\prime} 
    +\epsilon 
    \left[ 2 r^2 \Phi^{\prime \prime} +r\Phi^{\prime} \left(2 r \Phi^{\prime}-1\right) -2 \right]
    \right\} 
    \, , \nonumber \\
    R_{\theta \theta} 
    &=
    \, \frac{1}{2r} 
    \left[r\epsilon^{\prime} + 2 r (\epsilon-r) \Phi^{\prime} + \epsilon\right] 
    \, , \nonumber \\
    R_{\phi t} 
    &=
    \frac{r^3}{2} \sin ^4\theta  \e^{-2 \Phi} \omega (r-\epsilon) (\omega^{\prime})^{2} 
    \nonumber \\
    & \quad \ 
    - \frac{1}{4 r} \sin^2\theta 
    \left\{ 
    r \left[  
    r\omega ^{\prime} \left(2 r \Phi^{\prime}+\epsilon^{\prime}-8\right)-2 r^2 \omega^{\prime \prime}
    +2 \omega \left(\epsilon^{\prime}-2 r \Phi^{\prime}\right) 
    \right] 
    \right. \nonumber \\
    &\qquad \qquad \qquad \quad \left. 
    +\epsilon 
    \left[ 
    r \omega^{\prime} \left(7-2 r \Phi^{\prime}\right)  
    +2 r^2 \omega^{\prime \prime}
    +2\omega \left(2 r \Phi^{\prime}+1\right) 
    \right] 
    \right\} 
    \, , \nonumber \\
    R_{\phi \phi} 
    &=
    \frac{1}{2 r} \sin^2\theta 
    \left[ 
    r^4 \sin^2\theta \, \e^{-2 \Phi} (\epsilon-r) (\omega^{\prime})^{2}
    +r \epsilon^{\prime} +2 r (\epsilon-r) \Phi^{\prime} +\epsilon 
    \right]
    \, , \nonumber \\
    R 
    &=
    \frac{1}{r^2} 
    \left\{ 
    \epsilon^{\prime} \left(r \Phi^{\prime}+2\right) 
    +(3 \epsilon-4 r) \Phi^{\prime} -2 r (r-\epsilon) \left[\Phi^{\prime \prime}+(\Phi^{\prime})^2\right] 
    \right\} 
    +\frac{r}{2} \sin^2\theta \ \e^{-2 \Phi} (r-\epsilon) (\omega^{\prime})^2 
    \, .
\end{align}
By using the expressions of the Ricci tensor in Eqs.~\eqref{Ricci} and \eqref{Eqs5}, 
we can straightforwardly construct a non-linear $\sigma$ model realising the wormhole geometry described in Eq.~\eqref{DMWH1}. 

As in the previous sections, we consider the energy density, pressure and the energy flux in this wormhole geometry. 
The non-vanishing components of the Einstein tensor have the following forms:
\begin{align}
\label{EinsteinBBB}
    G_{t t} 
    &=
    \frac{1}{4 r^2} 
    \left[
    3 r^5 \sin ^4 \theta \, \e^{-2 \Phi} \omega^2 (\epsilon-r) (\omega^{\prime})^{2} 
    +  4 \e^{2 \Phi} \epsilon^{\prime} 
    \right]
    \nonumber \\
    & \quad \
    + \frac{1}{4 r} \sin^2\theta 
    \left\{ 
    r \left[ 
    -r^2 (\omega^{\prime})^{2} 
    +2 \omega^2 \left( 2 r^2 \Phi^{\prime \prime}-\left(r \Phi^{\prime}+1\right) \left(\epsilon^{\prime}-2 r \Phi^{\prime}\right) 
    \right) 
    + 2 r \omega \left( \omega^{\prime} \left(2 r \Phi^{\prime}+\epsilon^{\prime}-8\right)-2 r \omega^{\prime \prime} \right) 
    \right] 
    \right. \nonumber \\
    & \qquad \qquad \qquad \quad
    \left. 
    +\epsilon \left[ 
    r^2 (\omega^{\prime})^{2}
    -2 \omega^2 \left( 2 r^2 \Phi^{\prime \prime}+2 r^2 (\Phi^{\prime})^{2} + r\Phi^{\prime}-1\right) 
    +2 r \omega \left( \left(7-2 r \Phi^{\prime}\right) \omega^{\prime} +2 r \omega^{\prime \prime} \right) 
    \right] 
    \right\} 
    \, , \nonumber \\
    G_{r r} 
    &= 
    \frac{r^2}{4} \sin^2\theta \, \e^{-2 \Phi} (\omega^{\prime})^{2}
    + \frac{1}{r^2} \left( 2 r \Phi^{\prime}+\frac{r}{\epsilon-r}+1 \right) \, , \nonumber \\
    G_{\theta \theta} 
    &=
    \frac{r^3}{4} \sin^2\theta \, \e^{-2 \Phi} (\epsilon-r) (\omega^{\prime})^{2}
    +\frac{1}{2r} \left(r \Phi^{\prime}+1\right) \left[-r \epsilon^{\prime}+2 r (r-\epsilon) \Phi^{\prime}+\epsilon\right] 
    +r (r-\epsilon) \Phi^{\prime \prime} 
    \, , \nonumber \\
    G_{\phi t} 
    &=
    \frac{1}{4 r} \sin^2\theta 
    \left\{ 
    3 r^4 \sin^2\theta  \e^{-2 \Phi} \omega (r-\epsilon) (\omega^{\prime})^{2} 
    \right. \nonumber \\
    &\qquad \qquad \qquad 
    -r \left[ 
    4 r^2 \omega \Phi^{\prime \prime} 
    + r\omega^{\prime} \left(2 r \Phi^{\prime}+\epsilon^{\prime}-8\right) -2 r^2 \omega^{\prime \prime}
    -2 \omega \left(r \Phi^{\prime}+1\right) \left(\epsilon^{\prime}-2 r \Phi^{\prime}\right) 
    \right] 
    \nonumber \\
    &\qquad \qquad \qquad 
    \left. 
    +\epsilon 
    \left[ 
    r \omega^{\prime} \left(2 r \Phi^{\prime}-7\right) -2 r^2 \omega^{\prime \prime} 
    + 2 \omega 
    \left( 2 r^2 \Phi^{\prime \prime} +2 r^2 (\Phi^{\prime})^{2} + r\Phi^{\prime} -1 \right) 
    \right] 
    \right\} \, , \nonumber \\
    G_{\phi \phi} 
    &=
    \frac{1}{4 r} \sin^2\theta 
    \left\{ 
    3 r^4 \sin^2\theta  \e^{-2 \Phi} (\epsilon-r) (\omega^{\prime})^{2}
    +4 r^2 (r-\epsilon) \Phi^{\prime \prime} 
    +2 \left(r \Phi^{\prime}+1\right) 
    \left[-r \epsilon^{\prime}+2 r (r-\epsilon) \Phi^{\prime}+\epsilon \right] 
    \right\} 
    \, . 
\end{align}
For the four-velocity $u_\mu$ of the perfect fluid, the energy-momentum tensor has the following form, 
\begin{align}
\label{EMtensor}
    T_{\mu\nu} = (\rho + p) u_\mu u_\nu + p g_{\mu\nu} 
    \, .
\end{align}
The Einstein equation, $G_{\mu\nu} = \kappa^2 T_{\mu\nu}$, indicates that $u_\phi$ must be non-vanishing, in addition to $u_t$, since $G_{\phi t}$ does not vanish in Eq.~\eqref{EinsteinBBB}.
From the normalisation condition $u_\mu u^\mu = -1$, we obtain
\begin{align}
\label{4vctr}
     - \e^{-2\Phi(r)} {u_t}^2 
     + \left( \frac{\csc^2\theta}{r^2} - \e^{-2\Phi} \omega(r)^2 \right) {u_\phi}^2 
     - 2 \e^{-2\Phi(r)} \omega(r) u_t u_\phi 
     = 
     -1 
     \, .
\end{align}
If the fluid were genuine DM with vanishing pressure ($p=0$), the form of the energy–momentum tensor in Eq.~\eqref{EMtensor} and the Einstein equation $G_{\mu\nu} = \kappa^2 T_{\mu\nu}$ would require $G_{rr}$ and $G_{\theta\theta}$ to vanish.
However, these components remain non-vanishing as long as $\omega$ is not constant.
Therefore, the corresponding matter cannot represent cold DM, although the wormhole geometry in Eq.~\eqref{DMWH1} can still be realised within the four–scalar model.

We should note that any other kinds of wormhole geometries, based on other DM profiles or anything else, can also be stably realised by using the four-scalar non-linear $\sigma$ model given in \eqref{Eq. acg1} - \eqref{Eq. acg3}. 
We do not need to use any real DM or compensate the real DM by using the four-scalar model.  
Our analysis merely highlights a limitation of the proposed DM halo scenario~\cite{Errehymy:2025vvs} within the four-scalar non-linear  $\sigma$ model, and the absence of a pressureless DM component in this context can be viewed as a constraint on that specific application.

\section{Summary and Discussions}
\label{SecX}

In this work, we have investigated wormhole spacetimes embedded in an expanding universe.
We have shown that in some cases, the energy conditions can be satisfied owing to the expansion of the universe,
since the energy density and pressure driving the expansion modify the total energy density and pressure.
We have found that all the energy conditions can be satisfied when the radius of the wormhole throat is comparable to or larger than the Hubble radius.

When the EoS parameter satisfies $w \geq - 1/3$, all the energy conditions do not seem to be violated at least in the late universe. 
This occurs because the observable universe is contained within the wormhole.
At late time, the throat radius increases continuously due to cosmic expansion,
and the universe is eventually “swallowed” by the wormhole.
For $-1 < w < - 1/3$, the energy conditions are generally violated,
but all of them can be recovered at late times except for the SEC, which remains violated.
A similar tendency is found for $w=-1$, where all energy conditions except the SEC are restored asymptotically.
In the phantom regime ($w<-1$), all energy conditions are violated, as in the standard phantom universe.

We have also examined a more intriguing class of wormhole geometries.
In these spacetimes, while our universe encounters a finite future singularity,
another universe connected through the wormhole remains nonsingular.
The singularity appears on a space-like hypersurface, except near the wormhole throat,
where it becomes time-like.
Thus, there exist time-like trajectories along which a particle can travel to the other universe before the singularity
and return to our universe after the singularity through the wormhole.

Furthermore, we have examined wormhole geometries motivated by DM halo profiles proposed in Ref.~\cite{Errehymy:2025vvs}.
We have shown that in these geometries, the pressure cannot vanish.
Hence, the matter source cannot correspond to ordinary cold DM, but rather to a form of exotic matter, which can nevertheless be realised within the four-scalar framework.

In Ref.~\cite{Simpson:2018tsi}, a spacetime interpolating between the Schwarzschild black hole and the Morris–Thorne wormhole was proposed (see also \cite{Bronnikov:2021uta}),
\begin{align}
\label{SV}
    ds^2 
    = 
    - \left( 1 - \frac{2m}{\sqrt{r^2 + a^2}}\right) dt^2 
    + \left( 1 - \frac{2m}{\sqrt{r^2 + a^2}}\right)^{-1} dr^2 
    + \left( r^2 + a^2 \right) \left( d\theta^2 + \sin^2\theta d\phi^2 \right) 
    \, .
\end{align} 
Here $r$ ranges from $-\infty$ to $+\infty$.
For $r \gg a$, the geometry reduces to the Schwarzschild black hole,
\begin{align}
\label{SVB}
    ds^2 
    \approx 
    - \left( 1 - \frac{2m}{r}\right) dt^2 
    + \left( 1 - \frac{2m}{r}\right)^{-1} dr^2 
    + r^2 \left( d\theta^2 + \sin^2\theta d\phi^2 \right) 
    \, .
\end{align} 
and $m$ represents the ADM mass.
Thus, in the limit $a \to 0$, the metric recovers the Schwarzschild geometry.
When $0 \le a \le 2m$, horizons appear at
\begin{align}
\label{SVhorizon}
    r = \pm \sqrt{4m^2 - a^2}
    \, .
\end{align}
If $a > 2m$, the horizons disappear, and the two asymptotically flat regions ($r>0$ and $r<0$)
are smoothly connected through the throat of radius $a$ at $r=0$.
For $0 < a < 2m$, horizons exist but no singularity occurs;
the spacetime thus describes a regular black hole containing a wormhole inside the two horizons.
The horizon at negative $r$, i.e., $r = - \sqrt{4m^2 - a^2}$, may correspond to a white hole.

We would like to consider solutions in which the spacetime dynamically evolves
from a wormhole to a black hole by allowing $a$ to depend on time, $a=a(t)$.
However, such a dynamical transition introduces a curvature singularity at the former horizon.
Indeed, the scalar curvature is given by
\begin{align}
\label{scalarcurv}
    R 
    &=
    \frac{2}{\left(a(t)^2+r^2\right)^{5/2} \left(\sqrt{a(t)^2+r^2}-2 m\right)^3} 
    \left\{ a(t)^5 \ddot a(t) \left(-9 m \sqrt{a(t)^2+r^2}+10 m^2+6 r^2\right) \right. 
    \nonumber \\
    & \qquad
    +2 r^2 a(t)^3 \ddot a(t) \left(-9 m \sqrt{a(t)^2+r^2}+10 m^2+3 r^2\right) 
    +r^4 a(t) \ddot a(t) \left(-9 m \sqrt{a(t)^2+r^2}+10 m^2+2 r^2\right) 
    \nonumber \\
    & \qquad
    +2 a(t)^7 \ddot a(t)+a(t)^4 \left[ \dot a(t)^2 \left(-6 m \sqrt{a(t)^2+r^2}+12 m^2+4 r^2\right) 
    +9 m \sqrt{a(t)^2+r^2}-30 m^2-2 r^2 \right] 
    \nonumber \\
    & \qquad
    +r^4 \dot a(t)^2 \left(-9 m \sqrt{a(t)^2+r^2}+10 m^2+2 r^2\right)
    \nonumber \\
    & \qquad
    -a(t)^2 \left[ r^2 \dot a(t)^2 \left(15 m \sqrt{a(t)^2+r^2}-22 m^2-5 r^2\right) 
    -44 m^3 \sqrt{a(t)^2+r^2} \right. 
    \nonumber \\
    & \qquad
    \left. \left. -9 m r^2 \sqrt{a(t)^2+r^2} 
    +24 m^4+30 m^2 r^2+r^4 \right] 
    +a(t)^6 \left(\dot a(t)^2-1\right) \right\} 
    \, , 
\end{align}
which diverges at the horizon defined by Eq.~(\ref{SVhorizon}).
The resulting singularity is naked.
In general, we can show that when the horizon radius evolves dynamically,
the horizon tends to become a curvature singularity.

Finally, we make comments on thermodynamical aspects.
Since wormholes do not possess horizons, they do not emit Hawking radiation~\cite{Hawking:1975vcx}, unlike black holes.
Consequently, the notion of temperature is ill-defined, and a consistent thermodynamical description is challenging.
Several approaches have been proposed to define the entropy of wormholes, such as through entanglement entropy~\cite{Maldacena:2013xja} or the minimal area of the throat~\cite{Tomikawa:2015swa}, but no universally accepted definition has yet been established.
In the model~\eqref{SV}, the horizon emerges for $a^2 < 4m^2$, allowing one to discuss thermodynamic properties.
When $a^2 > 4m^2$, the horizon disappears, but it may be possible to define an entropy by analytic continuation.
However, as mentioned above, constructing a model in which the horizon radius dynamically changes without encountering singularities remains a difficult task.

\acknowledgments

T.K. is supported by the National Science Foundation of China (No.~12403003), and the National Key R\&D Program of China (No.~2021YFA0718500).



\begin{thebibliography}{99}

\bibitem{Misner:1960zz}
C.~W.~Misner,
Phys. Rev. \textbf{118} (1960), 1110-1111
doi:10.1103/PhysRev.118.1110

\bibitem{Wheeler:1957mu}
J.~A.~Wheeler,
Annals Phys. \textbf{2} (1957), 604-614
doi:10.1016/0003-4916(57)90050-7

\bibitem{Morris:1988cz}
M.~S.~Morris and K.~S.~Thorne,
Am. J. Phys. \textbf{56} (1988), 395-412
doi:10.1119/1.15620

\bibitem{Bronnikov:2002rn}
K.~A.~Bronnikov and S.~W.~Kim,
Phys. Rev. D \textbf{67} (2003), 064027
doi:10.1103/PhysRevD.67.064027
[arXiv:gr-qc/0212112 [gr-qc]].


\bibitem{Maeda:2009tk}
H.~Maeda, T.~Harada and B.~J.~Carr,
Phys. Rev. D \textbf{79} (2009), 044034
doi:10.1103/PhysRevD.79.044034
[arXiv:0901.1153 [gr-qc]].

\bibitem{Capozziello:2022zoz}
S.~Capozziello and N.~Godani,
Phys. Lett. B \textbf{835} (2022), 137572
doi:10.1016/j.physletb.2022.137572
[arXiv:2211.06481 [gr-qc]].

\bibitem{Godani:2024rqf}
N.~Godani,
Int. J. Geom. Meth. Mod. Phys. \textbf{21} (2024) no.13, 2450227
doi:10.1142/S021988782450227X




\bibitem{Bronnikov:2023saq}
K.~A.~Bronnikov, P.~E.~Kashargin and S.~V.~Sushkov,
Universe \textbf{9} (2023), 465
doi:10.3390/universe9110465
[arXiv:2309.03166 [gr-qc]].

\bibitem{Nojiri:2023dvf}
S.~Nojiri and G.~G.~L.~Nashed,
Phys. Rev. D \textbf{108} (2023) no.12, 124049
doi:10.1103/PhysRevD.108.124049
[arXiv:2309.12379 [hep-th]].

\bibitem{Elizalde:2023rds}
E.~Elizalde, S.~Nojiri, S.~D.~Odintsov and V.~K.~Oikonomou,
Phys. Dark Univ. \textbf{45} (2024), 101536
doi:10.1016/j.dark.2024.101536
[arXiv:2312.02889 [gr-qc]].


\bibitem{Nashed:2024jqw}
G.~G.~L.~Nashed and S.~Nojiri,
Phys. Dark Univ. \textbf{46} (2024), 101655
doi:10.1016/j.dark.2024.101655
[arXiv:2402.12860 [gr-qc]].

\bibitem{Katsuragawa:2024bwm}
T.~Katsuragawa, S.~Nojiri and S.~D.~Odintsov,
Phys. Rev. D \textbf{110} (2024) no.6, 064014
doi:10.1103/PhysRevD.110.064014
[arXiv:2406.18368 [gr-qc]].


\bibitem{Nojiri:2025heh}
S.~Nojiri and S.~D.~Odintsov,
Phys. Dark Univ. \textbf{50} (2025), 102104
doi:10.1016/j.dark.2025.102104
[arXiv:2508.07524 [gr-qc]].


\bibitem{deHaro:2023lbq}
J.~de Haro, S.~Nojiri, S.~D.~Odintsov, V.~K.~Oikonomou and S.~Pan,
Phys. Rept. \textbf{1034} (2023), 1-114
doi:10.1016/j.physrep.2023.09.003
[arXiv:2309.07465 [gr-qc]].

\bibitem{Navarro:1995iw}
J.~F.~Navarro, C.~S.~Frenk and S.~D.~M.~White,
Astrophys. J. \textbf{462} (1996), 563-575
doi:10.1086/177173
[arXiv:astro-ph/9508025 [astro-ph]].

\bibitem{Navarro:1996gj}
J.~F.~Navarro, C.~S.~Frenk and S.~D.~M.~White,
Astrophys. J. \textbf{490} (1997), 493-508
doi:10.1086/304888
[arXiv:astro-ph/9611107 [astro-ph]].

\bibitem{Errehymy:2025vvs}
A.~Errehymy, A.~Guvendi, S.~G.~Dogan and O.~Mustafa,
[arXiv:2509.16739 [gr-qc]].

\bibitem{Boehmer:2007um}
C.~G.~Boehmer and T.~Harko,
JCAP \textbf{06} (2007), 025
doi:10.1088/1475-7516/2007/06/025
[arXiv:0705.4158 [astro-ph]].

\bibitem{Begeman:1991iy}
K.~G.~Begeman, A.~H.~Broeils and R.~H.~Sanders,
Mon. Not. Roy. Astron. Soc. \textbf{249} (1991), 523
doi:10.1093/mnras/249.3.523

\bibitem{Nojiri:2020blr}
S.~Nojiri, S.~D.~Odintsov and V.~Faraoni,
Phys. Rev. D \textbf{103} (2021) no.4, 044055
doi:10.1103/PhysRevD.103.044055
[arXiv:2010.11790 [gr-qc]].

\bibitem{Alencar:2025nik}
G.~Alencar, R.~D{\'a}rlla, S.~Nojiri, S.~D.~Odintsov and D.~S{\'a}ez-Chill{\'o}n G{\'o}mez,
[arXiv:2508.05536 [gr-qc]].

\bibitem{Nojiri:2005sx}
S.~Nojiri, S.~D.~Odintsov and S.~Tsujikawa,
Phys. Rev. D \textbf{71} (2005), 063004
doi:10.1103/PhysRevD.71.063004
[arXiv:hep-th/0501025 [hep-th]].

\bibitem{Dabrowski:2009kg}
M.~P.~Dabrowski and T.~Denkieiwcz,
Phys. Rev. D \textbf{79} (2009), 063521
doi:10.1103/PhysRevD.79.063521
[arXiv:0902.3107 [gr-qc]].

\bibitem{Caldwell:2003vq}
R.~R.~Caldwell, M.~Kamionkowski and N.~N.~Weinberg,
Phys. Rev. Lett. \textbf{91} (2003), 071301
doi:10.1103/PhysRevLett.91.071301
[arXiv:astro-ph/0302506 [astro-ph]].

\bibitem{Caldwell:1999ew}
R.~R.~Caldwell,
Phys. Lett. B \textbf{545} (2002), 23-29
doi:10.1016/S0370-2693(02)02589-3
[arXiv:astro-ph/9908168 [astro-ph]].

\bibitem{Barrow:2004xh}
J.~D.~Barrow,
Class. Quant. Grav. \textbf{21} (2004), L79-L82
doi:10.1088/0264-9381/21/11/L03
[arXiv:gr-qc/0403084 [gr-qc]].

\bibitem{Nojiri:2004pf}
S.~Nojiri and S.~D.~Odintsov,
Phys. Rev. D \textbf{70} (2004), 103522
doi:10.1103/PhysRevD.70.103522
[arXiv:hep-th/0408170 [hep-th]].

\bibitem{Bouhmadi-Lopez:2006fwq}
M.~Bouhmadi-Lopez, P.~F.~Gonzalez-Diaz and P.~Martin-Moruno,
Phys. Lett. B \textbf{659} (2008), 1-5
doi:10.1016/j.physletb.2007.10.079
[arXiv:gr-qc/0612135 [gr-qc]].

\bibitem{Capolupo:2024ckb}
A.~Capolupo, S.~Capozziello, G.~Pisacane and A.~Quaranta,
Phys. Dark Univ. \textbf{48} (2025), 101894
doi:10.1016/j.dark.2025.101894
[arXiv:2411.17319 [hep-ph]].

\bibitem{Rahaman:2023tkm}
F.~Rahaman, B.~Samanta, N.~Sarkar, B.~Raychaudhuri and B.~Sen,
Eur. Phys. J. C \textbf{83} (2023) no.5, 395
doi:10.1140/epjc/s10052-023-11456-4

\bibitem{Simpson:2018tsi}
A.~Simpson and M.~Visser,
JCAP \textbf{02} (2019), 042
doi:10.1088/1475-7516/2019/02/042
[arXiv:1812.07114 [gr-qc]].

\bibitem{Bronnikov:2021uta}
K.~A.~Bronnikov and R.~K.~Walia,
Phys. Rev. D \textbf{105} (2022) no.4, 044039
doi:10.1103/PhysRevD.105.044039
[arXiv:2112.13198 [gr-qc]].

\bibitem{Hawking:1975vcx}
S.~W.~Hawking,
Commun. Math. Phys. \textbf{43} (1975), 199-220
[erratum: Commun. Math. Phys. \textbf{46} (1976), 206]
doi:10.1007/BF02345020

\bibitem{Maldacena:2013xja}
J.~Maldacena and L.~Susskind,
Fortsch. Phys. \textbf{61} (2013), 781-811
doi:10.1002/prop.201300020
[arXiv:1306.0533 [hep-th]].

\bibitem{Tomikawa:2015swa}
Y.~Tomikawa, K.~Izumi and T.~Shiromizu,
Phys. Rev. D \textbf{91} (2015) no.10, 104008
doi:10.1103/PhysRevD.91.104008
[arXiv:1503.01926 [gr-qc]].










\end{thebibliography}
\end{document}